\documentclass{aa}  
\usepackage{graphicx}
\usepackage{amsmath} 
\usepackage{txfonts}
\usepackage[T1]{fontenc}
\usepackage{afterpage}
\usepackage{amsmath}
\usepackage{amssymb}
\usepackage{natbib} 
\bibpunct{(}{)}{;}{a}{}{,}
\newcommand{\Ms}{\ensuremath{M_\odot}}
\newcommand{\Zs}{\ensuremath{Z_\odot}}

\newcommand{\beqa}{\begin{eqnarray*}}
\newcommand{\eeqa}{\end{eqnarray*}}
\newcommand{\beqan}{\begin{eqnarray}}
\newcommand{\eeqan}{\end{eqnarray}}
\newcommand{\beq}{\begin{equation}}
\newcommand{\eeq}{\end{equation}}
\newcommand{\diff}{{\rm d}}
\newcommand{\drr}{\frac{\partial}{\partial r}}
\newcommand{\dtt}{\frac{\diff}{\diff t}}
\newcommand{\dr}[1]{\frac{\partial  #1}{\partial r}}

\newcommand{\lp}{ \left(}
\newcommand{\rp}{ \right)}
\newcommand{\lc}{ \left[}
\newcommand{\rc}{ \right]}

\usepackage{hyperref}
\hypersetup{colorlinks}
\usepackage{color}

\makeatother

\authorrunning{C.Charbonnel et al.}  \titlerunning{Impact of internal gravity waves on the rotation profile inside pre-main sequence low-mass stars} 

\begin{document}

\title{Impact of internal gravity waves on the rotation profile inside pre-main sequence low-mass stars}


\author{C. Charbonnel\inst{1,2} \and T.Decressin\inst{1} \and L.Amard\inst{1,3} \and A.Palacios\inst{3} \and S.Talon\inst{4}} 

\offprints{C.Charbonnel,\\ email: Corinne.Charbonnel@unige.ch}

\institute{Geneva Observatory, University of Geneva, Chemin des
  Maillettes 51, 1290 Versoix, Switzerland \and
              IRAP, CNRS UMR 5277, Universit\'e de Toulouse, 14, Av. E.Belin, 31400 Toulouse, France\and
              LUPM, Universit\'e Montpellier II, CNRS, UMR 5299, Place E. Bataillon, 34095, Montpellier, France \and
             Calcul Qu\'ebec, Universit\'e de Montr\'eal (DGTIC), C.P. 6128, succ. Centre-ville, Montr\'eal (Qu\'ebec) H3C 3J7
         \
}

    \date{Accepted for publication in A\&A (Section 7)} 

\abstract
  {} 
   {We study the impact of internal gravity waves (IGW), meridional circulation, shear turbulence, and stellar contraction on the internal rotation profile and surface velocity evolution of solar metallicity low-mass pre-main sequence stars.}
   {We compute a grid of rotating stellar evolution models with masses between 0.6 and 2.0~M$_{\odot}$ taking these processes into account for the transport of angular momentum, as soon as the radiative core appears and assuming no more disk-locking from that moment on. 
   IGW generation along the PMS is computed taking Reynolds-stress and buoyancy into account in the bulk of the stellar convective envelope and convective core (when present). 
   Redistribution of angular momentum within the radiative layers accounts for damping of prograde and retrograde IGW by thermal diffusivity and viscosity in corotation resonance. 
   }
   {Over the whole mass range considered, IGW are found to be efficiently generated by the convective envelope and to slow down the stellar core early on the PMS.
   In stars more massive than $\sim$ 1.6~M$_{\odot}$, IGW produced by the convective core also contribute to angular momentum redistribution close to the ZAMS.
   }
   {Overall, IGW are found to significantly change the internal rotation profile of PMS low-mass stars.}


\maketitle

\section{Introduction}
\label{section:introduction}

The evolution of the surface rotation of low-mass stars along the
pre-main sequence (hereafter PMS) follows a specific path as shown 
by the data, both rotation periods and $v \sin i$ measurements,
collected in young stellar clusters 
\citep[e.g.][ and references therein for a review]{IrwinBouvier2009}. 
The rotational properties of young stars appear to result from an intricate interplay between several physical processes that affect the angular
momentum gains, losses, and redistribution as the stars evolve along the
PMS towards the zero age main sequence (hereafter ZAMS). 
These mechanisms can be roughly divided into two classes. 
The first ones result from the connection of the stars to their environment (magnetic and dynamic coupling to a
circumstellar disk, accretion, jets, stellar and disk winds, etc.), and are particularly crucial during the T-Tauri phase, when star-disk interaction is observed and
expected to be strong \citep[see, for instance,][]{Shu1994,MattPudritz2005,ZanniFerreira2012}.
The broad variety of possible star-environment configurations may, in particular, explain part of the large dispersion in rotation rates of solar-type stars observed along the PMS and at the arrival on the ZAMS \citep[e.g., ][]{Staufferetal85,IrwinBouvier2009}.

The second class is related to stellar secular evolution and consists of the (magneto-) hydrodynamical transport 
mechanisms that contribute to redistributing angular momentum inside the stars themselves. 
In the present paper we focus on exploring these internal mechanisms once the disk has dissipated and the accretion process is over, which occurs after 3-10 Myr 
\citep[e.g.,][]{Haischetal01,Hartmann05,Hernandezetal08}, i.e., roughly at the time when a radiative core appears in the contracting PMS stars.
Our aim is to evaluate, in particular and for the first time, the 
interplay between internal gravity waves
(hereafter IGW), meridional circulation, turbulent shear, and stellar
contraction during the PMS, considering that IGW are one of the best candidate mechanisms
to explain the flat angular velocity profile inside the Sun as
revealed by helioseismology \citep{CharbonnelTalon2005Science}.
This work is also motivated by the results of  \citet[][ hereafter TC08]{TalonCharbonnel2008}, who showed that IGW are efficiently excited inside intermediate-mass PMS stars and 
suggested that waves should efficiently transport angular momentum during the PMS evolution, which should
affect the angular velocity profile at this phase and at the arrival on the ZAMS.

In \S\ref{section:formalism} we introduce the formalism for IGW excitation and for the transport of angular momentum through the various mechanisms considered. 
We describe in \S \ref{sec:models} the basic assumptions for the present grid of PMS models for low-mass (0.6 to 2~M$_{\odot}$), solar-metallicity stars.
In \S \ref{sec:IGWexcitation} we examine IGW generation by the convective envelope and the convective core (when present) 
along the PMS evolution for the whole mass range covered by the grid. 
In \S~\ref{evolutionrotprofile} we describe the impact of the various interacting transport mechanisms on the evolution of the internal rotation profile for the various stellar masses, and in \S~\ref{section:globalproperties} we briefly discuss their influence on the surface rotation velocity and lithium abundance, as well as on global stellar properties. 
Conclusions are presented in \S~\ref{sec:conclusion}.
 
\section{Formalism}
\label{section:formalism} 

We follow \citet[][ hereafter TC05]{TalonCharbonnel2005} for the treatment of both the excitation of IGW and the transport of angular momentum and chemicals 
by waves, meridional circulation, and shear turbulence in hydrodynamical stellar models. 
We, however, underline three main improvements over TC05 paper. First, we consider IGW generated both by the external and central convective regions (when present), while only those excited by the convective envelope were considered in our previous work. Second, the important variations in the stellar structure and of the IGW properties along the pre-main sequence (see \S~\ref{sec:IGWexcitation} and TC08)  
require that we then compute the wave spectra at each evolution time step, 
while the main sequence computation presented in TC05 was based on the wave spectrum of the stellar model on the ZAMS. Finally, we account here for both prograde and retrograde waves in the whole radiative interior, while only the latter ones were considered previously. 
We recall below the formalism (i.e.,  relevant equations and assumptions) that is included in the evolution code STAREVOL (see TC05, TC08, and \citealt{Mathisetal13}).

\subsection{IGW generation}
\label{subsection:excitation}
In this exploratory work we apply the \citet{GoldreichMurray1994} formalism as adapted by \citet{KQ1997} to calculate the spectrum of IGW excited by Reynolds stress and buoyancy in the bulk of convective regions \citep[see e.g.][]{ZahnTalon1997}. 
We do not consider the possible effects of IGWs generated by convective overshooting plumes, since no analytical prescription is available to describe this excitation mechanism (see details and discussion in TC05).  
However as we see in \S~\ref{subsubsection:generaltransport}, we multiply the IGW flux by a factor 2 in the transport equation in order to account for the recent results by \citet{LecoanetQuataert13}. 
We treat the waves by assuming that they are pure gravity waves (i.e., not modified by the Coriolis acceleration) that only feel the entrainment by differential rotation.

The kinetic energy flux per unit frequency is 
\begin{eqnarray}
{\cal F}_E \lp \ell, \omega \rp &=& \frac{\omega^2}{4\pi} \int dr\; \frac{\rho^2}{r^2}
   \left[\left(\frac{\partial \xi_r}{\partial r}\right)^2 +
   \ell(\ell+1)\left(\frac{\partial \xi_h}{\partial r}\right)^2 \right]  \nonumber \\
 && \times  \exp\left[ -h_\omega^2 \ell(\ell+1)/2r^2\right] \frac{v_c^3 L^4 }{1
  + (\omega \tau_L)^{15/2}}
\label{gold}
\end{eqnarray}
with $\xi_r$ and $[\ell(\ell+1)]^{1/2}\xi_h$ the radial and horizontal displacement wave functions normalized to unit energy flux at the
edge of the considered convection zone, $v_c$ the convective velocity,
$L=\alpha_{\rm MLT} H_P$ the radial size of an energy bearing turbulent eddy,
$\tau_L \approx L/v_c$ the characteristic convective time, $H_P$ the pressure scale height $P/\rho g$, and $h_\omega$ the radial size of the largest eddy at depth $r$ with characteristic frequency of $\omega$ or higher ($h_\omega = L \min\{1, (2\omega\tau_L)^{-3/2}\}$). 
The radial and horizontal wave numbers  (respectively $k_r$ and $k_h$) are related by
\beq
k_r^2 = \lp \frac{N^2}{\omega^2} -1 \rp k_h^2 = 
\lp \frac{N^2}{\omega^2} -1 \rp \frac{\ell \lp \ell +1 \rp}{r^2} \label{kradial}
\eeq
where $N^2$ is the Brunt-V\"ais\"al\"a frequency\footnote{The Brunt-V\"ais\"al\"a - or buoyancy - frequency $N$ is given by
$N^2 = N_T^2+N_\mu^2 = \frac{g}{H_P}\lp \delta(\nabla_\text{ad}-\nabla)+\varphi\nabla_\mu\rp$, with $\delta = −(\partial \ln \rho/\partial \ln T )_{P,µ}$, $\varphi =
(\partial \ln \rho/\partial \ln \mu)_{P,T}$,  $\nabla$ the logarithmic temperature gradient, and $\nabla_\mu$ the mean molecular weight gradient.} .

At the considered convective edge (located at radius $r_{cz}$), the mean flux of angular momentum carried by a monochromatic wave of spherical order $\ell$ and local (i.e., emission) frequency $\omega$ ($m$ being the azimutal order), i.e., the  momentum flux per unit frequency, 
is related to  the kinetic energy flux by 
\begin{equation}
{{\cal F}_{J,{r_{cz}}}}\lp m, \ell, \omega \rp = \frac{2m}{\omega} {\cal F}_E\lp \ell, \omega \rp  
\end{equation}   
 \citep{GoldreichNicholson1989,ZahnTalon1997}. 
 The so-called angular momentum luminosity at the considered convective edge (envelope or core) is obtained after horizontal
integration 
\beq
{{\cal L}_J}_{\ell, m} \lp r_{\rm cz} \rp = 4 \pi r_{cz}^2 {\cal F}_{J,r_{cz}}.
\label{eq:angmomluminosity}
\eeq

\subsection{IGW damping}
\label{subsection:transportIGW}

Deposition of angular momentum (positive or negative) within the radiative layers occurs
at the depth where individual monochromatic waves are eventually damped by thermal diffusivity and viscosity in corotation resonance \citep{GoldreichNicholson1989,Schatzman1993,ZahnTalon1997}.
In the present study the local momentum luminosity at a given radius $r$ within the radiative region accounts for prograde and retrograde waves  (i.e., with respectively positive and negative $m$ values) generated by both the convective envelope and the convective core (if present), i.e., 
\beq
{\cal L}_J(r) = {\cal L}_{J\text{,env}}+{\cal L}_{J,\text{core}}\label{eq:envcore},\eeq
where each component is given by  
\beq  
\sum_{\sigma, \ell, m} {{\cal L}_J}_{\ell, m} \lp r_{\rm cz}\rp
\exp \lc -\tau(r, \sigma, \ell) \rc \label{locmomlum},
\eeq
where `${\rm cz}$' refers to the interface between the radiative region and the corresponding convection zone (i.e., envelope or core).

The local damping rate
 \beq
\tau(r, \sigma, \ell) = \lc \ell(\ell+1) \rc ^{3\over2} \int_r^{r_{cz}} 
\lp K_T + \nu_v \rp \; {N {N_T}^2 \over
\sigma^4}  \left({N^2 \over N^2 - \sigma^2}\right)^{1 \over 2} {\diff r
\over r^3} \label{optdepth}
\eeq
 takes the mean molecular weight stratification into account  \citep{GoldreichNicholson1989,Schatzman1993,ZahnTalon1997}, as well as 
 the thermal and the (vertical) turbulent viscosity  ($K_T$ and $\nu_v$ respectively). 
Here, $\sigma$ is the local Doppler-shifted frequency
\beq
\sigma(r) = \omega - m
\lc \Omega(r)-\Omega _{\rm cz} \rc \label{sigma}
\eeq
with $\omega$ the wave frequency in the reference frame of the corresponding emitting convection zone that rotates with the angular velocity $\Omega _{\rm cz}$. 

As can be seen from these expressions, angular momentum redistribution by IGW within the radiative region is dominated by low-frequency ($\sigma \ll N$), low-degree waves; 
indeed, those penetrate deeper, and their prograde and retrograde components 
experience strong differential damping, as required to produce a net momentum deposition.
In contrast, high-degree waves are damped closer to the convection zone (since damping $\propto \lc \ell(\ell+1) \rc ^{3\over2}$), and high-frequency waves experience less differential damping.

\subsection{Global transport of angular momentum by IGW, meridional circulation, and shear turbulence}
\label{subsection:transportglobal}

\subsubsection{General equations}
\label{subsubsection:generaltransport}

We assume solid-body rotation in the convective regions. 
In the stellar radiative regions, the evolution of angular momentum through advection by meridional circulation, diffusion by shear turbulence, and deposit or extraction by IGW follows the general expression below \citep[e.g.,][]{TalonZahn1998}:
\beqan
\rho \dtt \lc r^2 {\Omega}\rc &= &
\frac{1}{5 r^2} \drr \lc \rho r^4 \Omega U \rc 
+ \frac{1}{ r^2} \drr \lc \rho \nu_v r^4 \dr{\Omega} \rc 
\nonumber \\
&& - {2} \frac{3}{8\pi} \frac{1}{r^2} \drr{{\cal L}_J(r)},
\label{ev_omega}
\eeqan  
where $U$ is the radial meridional circulation velocity, $\nu_v$ the turbulent viscosity due to differential rotation, and $\rho$ the density.
We have added a factor 2 in the last term to account for the study by \citet{LecoanetQuataert13}, who predict the IGW flux due to turbulent convection to be a few to five times larger than in previous estimates by, e.g., \citet{GoldreichKumar90} and \citet{GoldreichMurray1994}. However as we see in \S~\ref{subsubsect:mershearwaves}, our conclusions are not sensitive to this multiplication factor.

Following \citet{DecressinMathis2009} and \citet{Mathisetal13} we integrate Eq.~\ref{ev_omega} over 
an  isobar enclosing the mass $m\left(r\right)$  to obtain the expression of the total flux (loss or gain) of angular momentum carried by the considered transport processes:
\begin{eqnarray}
\Gamma(m) = \frac{1}{4\pi}\frac{\rm d}{{\rm d}t}\left[\int_{0}^{m\left(r\right)}{r'}^2\overline{\Omega}\, {\rm d}m'\right]  \nonumber  = - F_{\rm MC}\left(r\right) - F_{\rm S}\left(r\right) + F_{\rm IGW}\left(r\right)
\label{fluxAM}
\end{eqnarray}
where the fluxes driven by  meridional circulation, vertical shear-induced turbulence,  and IGWs are, respectively, 
\begin{equation}
F_{\rm MC}\left(r\right)=-\frac{1}{5}\overline\rho r^4{\overline\Omega}U_{2}
\label{eq:Fmc}
\end{equation}
\begin{equation}
F_{\rm S}\left(r\right)= - \overline\rho r^4 \nu_{v}\partial_{r}{\overline\Omega}
\label{eq:FS}
\end{equation}
\begin{equation}
F_{\rm IGW}\left(r\right)= \frac{3}{8 \pi} {\cal L}_J(r).
\label{eq:FIGW}
\end{equation}

\subsubsection{Meridional circulation}
\label{subsubsection:meridionalcirculation}
As can be seen in Eq.~\ref{ev_omega} the transport of angular momentum through meridional circulation is treated as an advective process. 
As in our previous studies we apply the formalism developed by
\citet{Zahn1992}, \citet{MaederZahn1998} and \citet[][ see also \citealt{DecressinMathis2009}]{MathisZahn2004}.

\subsubsection{Shear-induced turbulence}
\label{subsubsection:turbulence}
Shear-induced  turbulence is assumed to be highly anisotropic. 
Following TC05 we assume that the turbulent diffusion coefficient equals turbulent viscosity and use the corresponding expression given by  \citet{TalonZahn1997}, i.e.,  
\beq
 D_v = \nu_v = \frac{8}{5} \frac {Ri_{\rm crit}  (r \diff
 \Omega/\diff r)^2}{N^{2}_{T}/(K+D_h)+N^{2}_{\mu}/D_h}
\label{Dv}
\eeq
that considers the weakening effect of thermal diffusivity ($K_T$) on the thermal
stratification and of horizontal turbulence ($D_h$, see below) on both the thermal and mean
molecular weight stratifications.

For the treatment of horizontal turbulent viscosity, we follow \citet{Zahn1992}, again as in TC05: 
\begin{eqnarray}
D_h = \nu_h = \frac{r}{C_h}\sqrt{\left|\frac{1}{3 \rho r}\frac{\diff (\rho r^2 U)}{\diff
    r}-\frac{U}{2}\frac{\diff \ln r^2\Omega}{\diff \ln r}\right|^2 + U^2}
\label{eq:Dh}
\end{eqnarray}
with $C_h=1$.

The influence of the prescriptions assumed for $D_v$ and $D_h$ will be investigated in a future paper.

\subsection{Transport of chemicals}
We treat the transport of chemical species in the radiative regions as a diffusive process through the combined action of meridional circulation and shear-induced turbulence  \citep{ChaboyerZahn1992}. 
The effective diffusion coefficient is written 
\beq
D_{\rm eff} = \frac{ \left| r U(r) \right|^2}{30\,D_h}
\label{eq:Deff}
\eeq
where $D_h$ is the horizontal component of the turbulent diffusivity (see Eq.~\ref{eq:Dh}). 

In the present study we neglect atomic diffusion, whose effects require much longer timescales to develop compared to the very short duration of the pre-main sequence phase. 
We also neglect possible wave-induced turbulence. 
Therefore the expression for the transport of chemicals (here, the mass fraction $X$ of the element $i$) in the stellar radiative region writes as \citep[see e.g.][]{MeynetMaeder2000}:
\begin{eqnarray}
  \left(\frac{{\rm d} X_i}{{\rm d}t}\right)_{M_r}= \frac{\partial}{\partial
  M_r}\left[(4\pi
  r^2\rho)^2\left(D_V+D_{\rm eff}\right) \frac{\partial X_i}{\partial M_r}\right] + \left(\frac{{\rm d}X_i}{{\rm d}t}\right)_{\rm nucl},
\label{eq:transportchemicals}
\end{eqnarray}
where $d M_{r}=4\pi{\overline\rho} r^2 d r$, and the last term accounts for nuclear destruction or production of the considered element.

\section{Stellar models}
\label{sec:models}

\subsection{Input physics and basic assumptions}
\label{subsec:inputphysics}

We focus on the pre-main sequence evolution of solar-metallicity stars in the mass range between 0.6 and 2.0~M$_{\odot}$. 
We adopt the solar composition of \citet{AsplundGrevesse2009}. 
Opacity tables are updated accordingly both at high and low temperature respectively from OPAL and Wichita websites\footnote{\url{http://adg.llnl.gov/Research/OPAL/opal.html}; \url{http://webs.wichita.edu/physics/opacity/}} \citep[see e.g.][]{IglesiasRogers1996,FergusonAlexander2005}.
The mixing length parameter $\alpha_\text{MLT} = 1.63$ is calibrated so that our
standard (i.e., non rotating) 1 \Ms{}, \Zs{} model fits the solar radius,
effective temperature, and luminosity at the age of the sun. 
Convection zone bounderies are defined by the Schwarzschild criterion, 
and we do not account for convective overshoot.

\begin{figure*}  
  \includegraphics[width=0.5\textwidth]{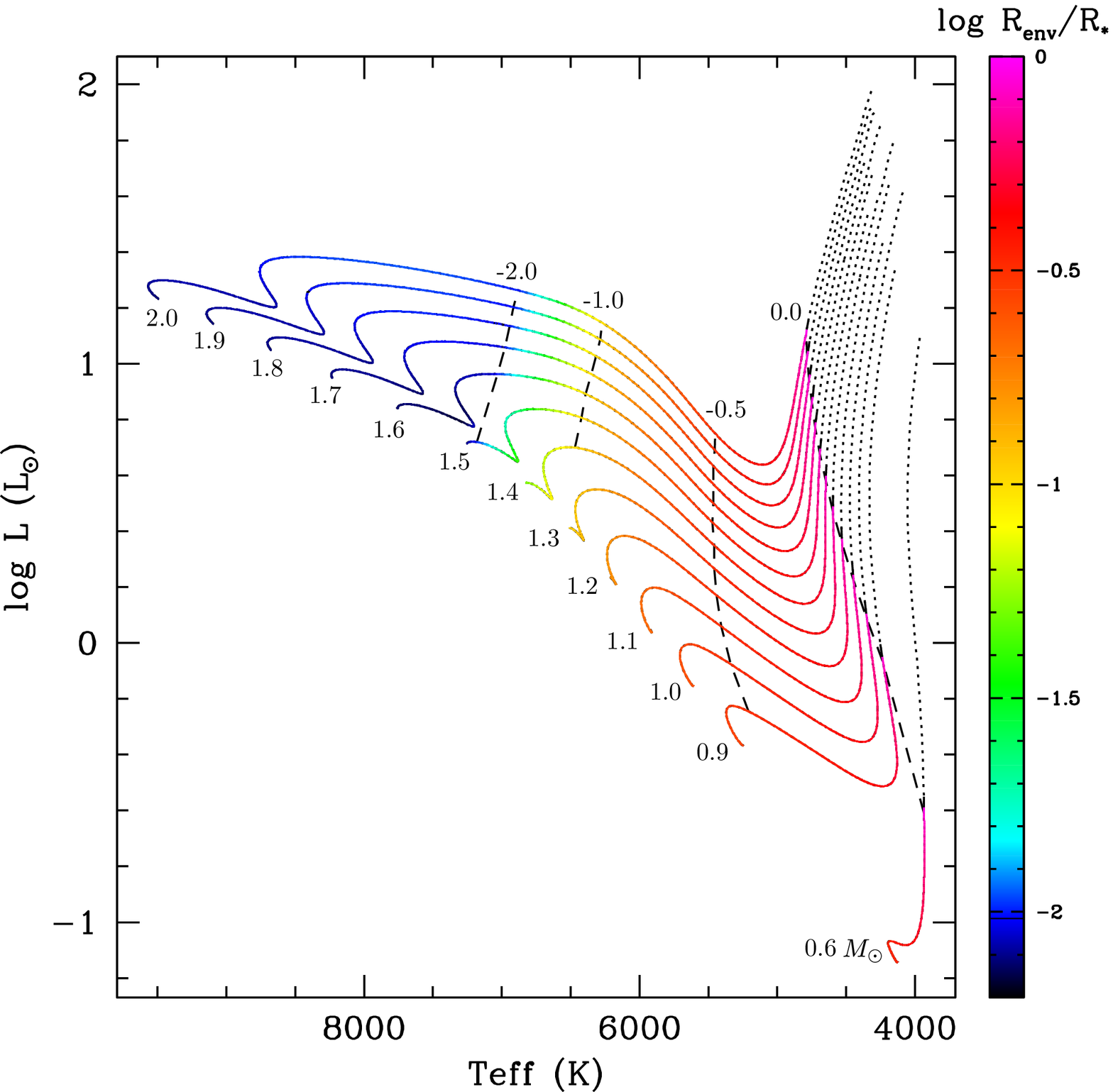}%
  \includegraphics[width=0.5\textwidth]{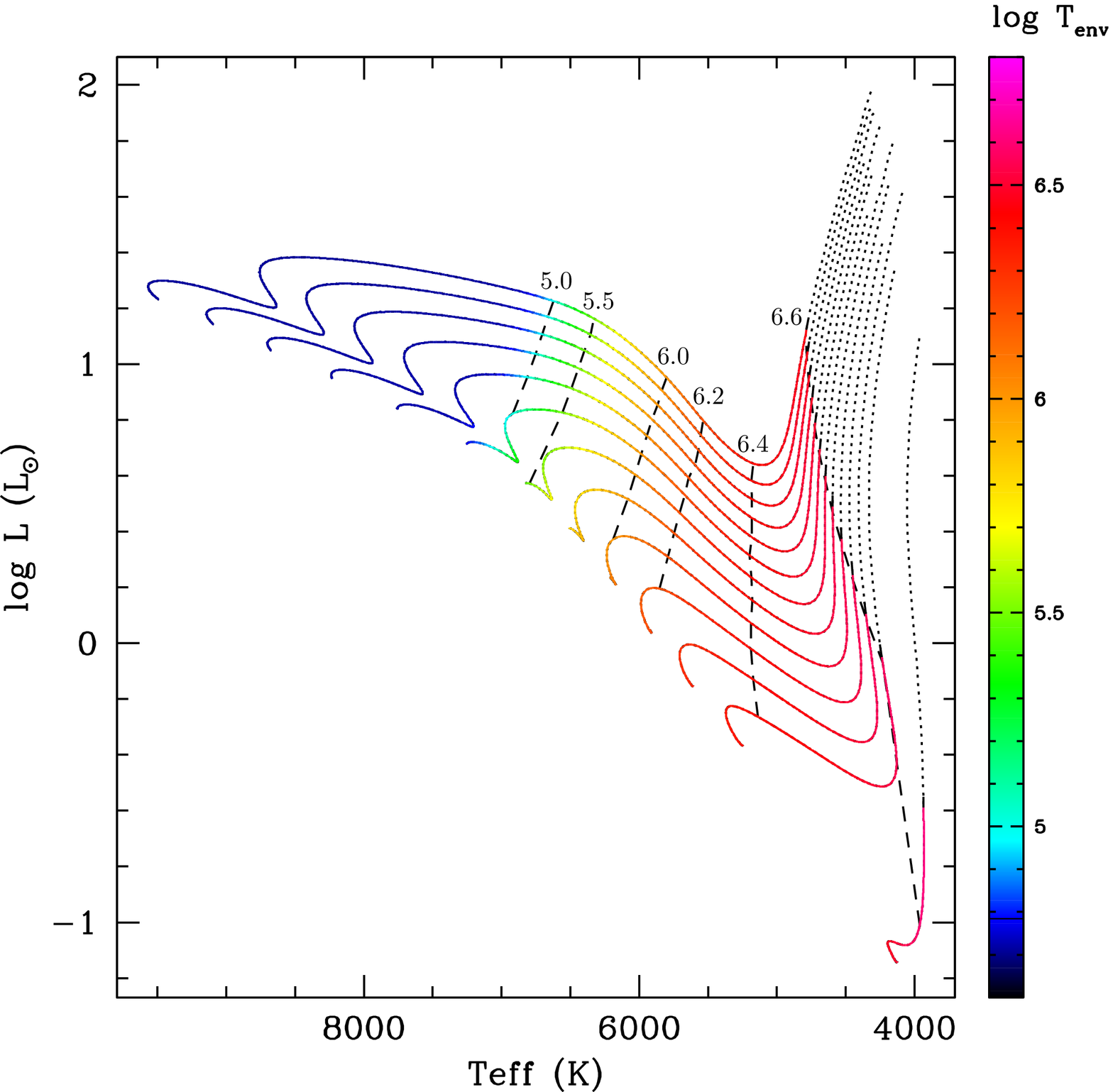}\\%
  \includegraphics[width=0.5\textwidth]{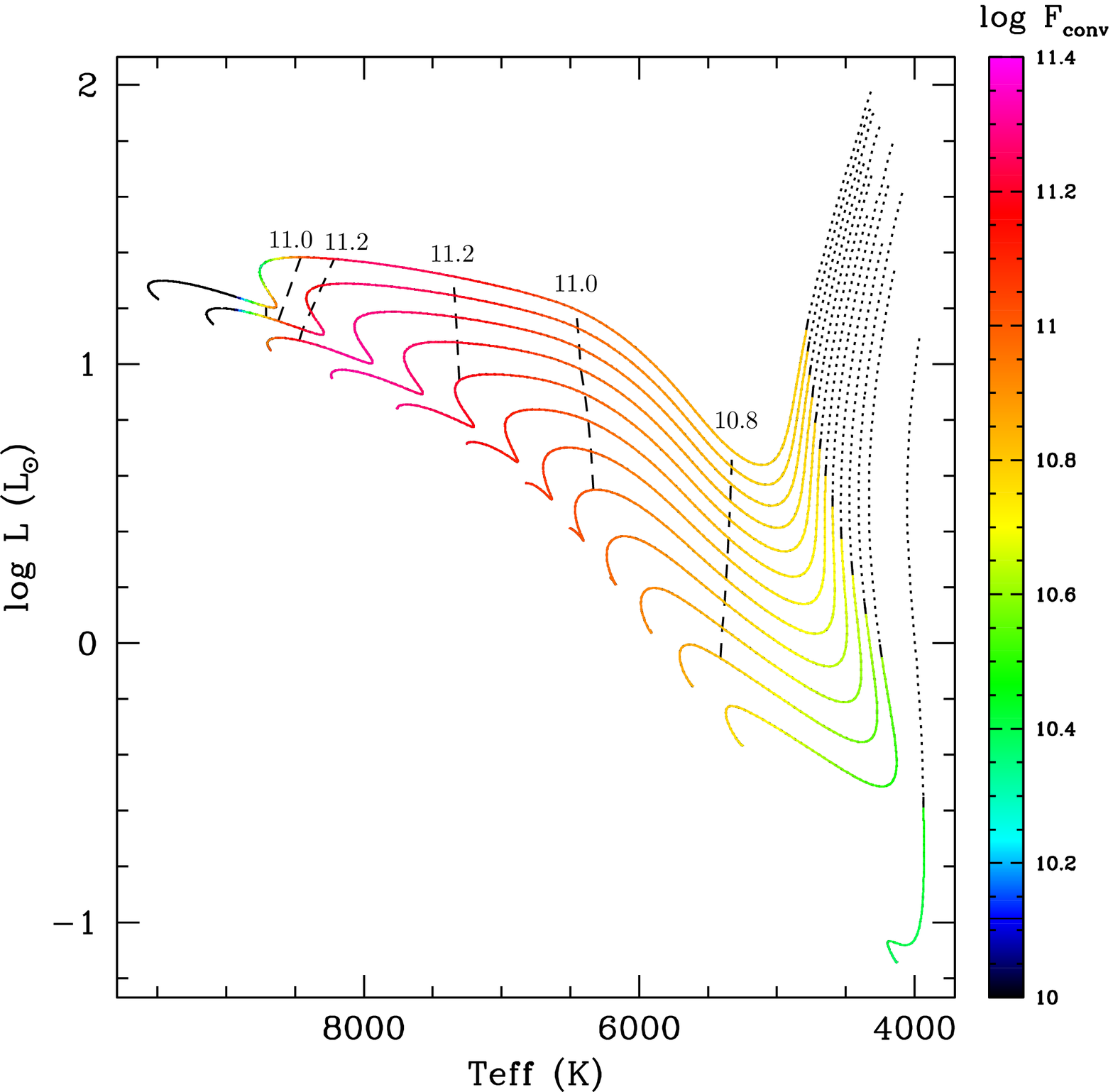}%
  \includegraphics[width=0.5\textwidth]{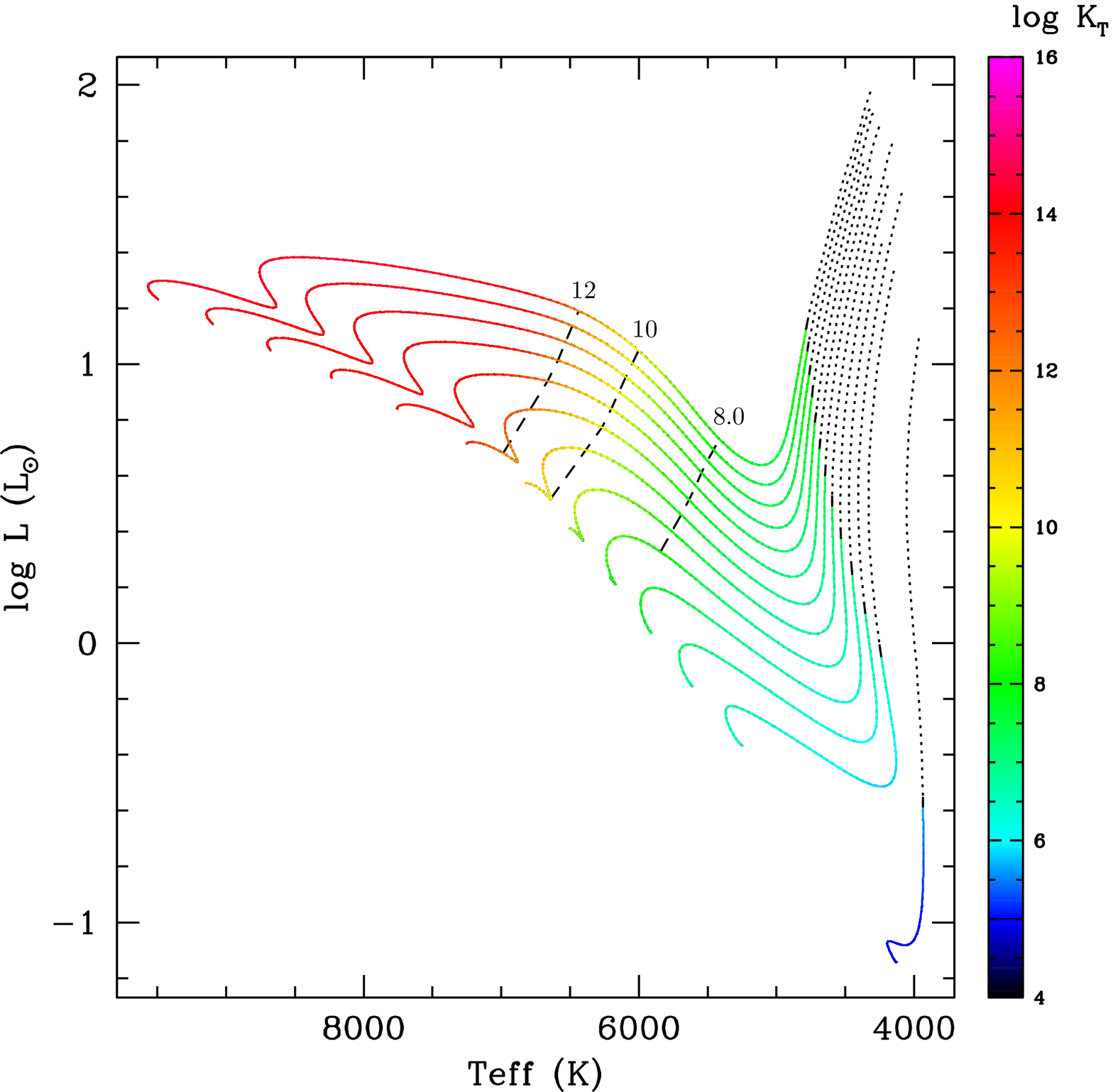}%
  \caption{PMS tracks in the Hertzsprung-Russel diagram for solar metallicity stars with initial masses between 0.6 and 2.0~\Ms{}
    (classical models are shown here) and properties of the external convective layers. 
    Colors indicate the radial extent of the convective envelope (top left panel), 
    the temperature at its bottom
    (top right panel), the maximal convective flux (bottom left panel), and
    the thermal diffusivity below the envelope (bottom right panel). 
    Dashed lines connect points with similar values for these quantities, and the colored axes are in cgs units.
    The dotted parts of the tracks correspond to the phase when the stars are still fully convective
    }
  \label{fig:surfdetails}
\end{figure*}

Computations are performed with the stellar evolution code STAREVOL \citep[see e.g. TC05, ][]{LagardeDecressin2012}.
Initial models are totally convective polytropic stars, with central temperature lower than 10$^6$~K (i.e., deuterium burning has not yet occurred). 
We follow the PMS evolution along the Hayashi track 
up to the arrival on the ZAMS that we define as the point where the ratio between central and surface hydrogen abundance reaches
0.998. 
The stellar mass is assumed to be constant during that phase (i.e., no accretion nor mass loss).
For each stellar mass we compute classical models (i.e., without any transport of angular momentum nor of chemicals) as well as rotating models with and without IGW. 
We neglect the hydrostatic effects of the centrifugal force in all our rotating models but two; we discuss the impact of this simplification in \S~\ref{evolutionrotprofile}.
The evolution tracks of the classical models in the Hertzsprung-Russel diagram are shown in Fig.\ref{fig:surfdetails}.

\begin{figure*}  
  \includegraphics[width=0.5\textwidth]{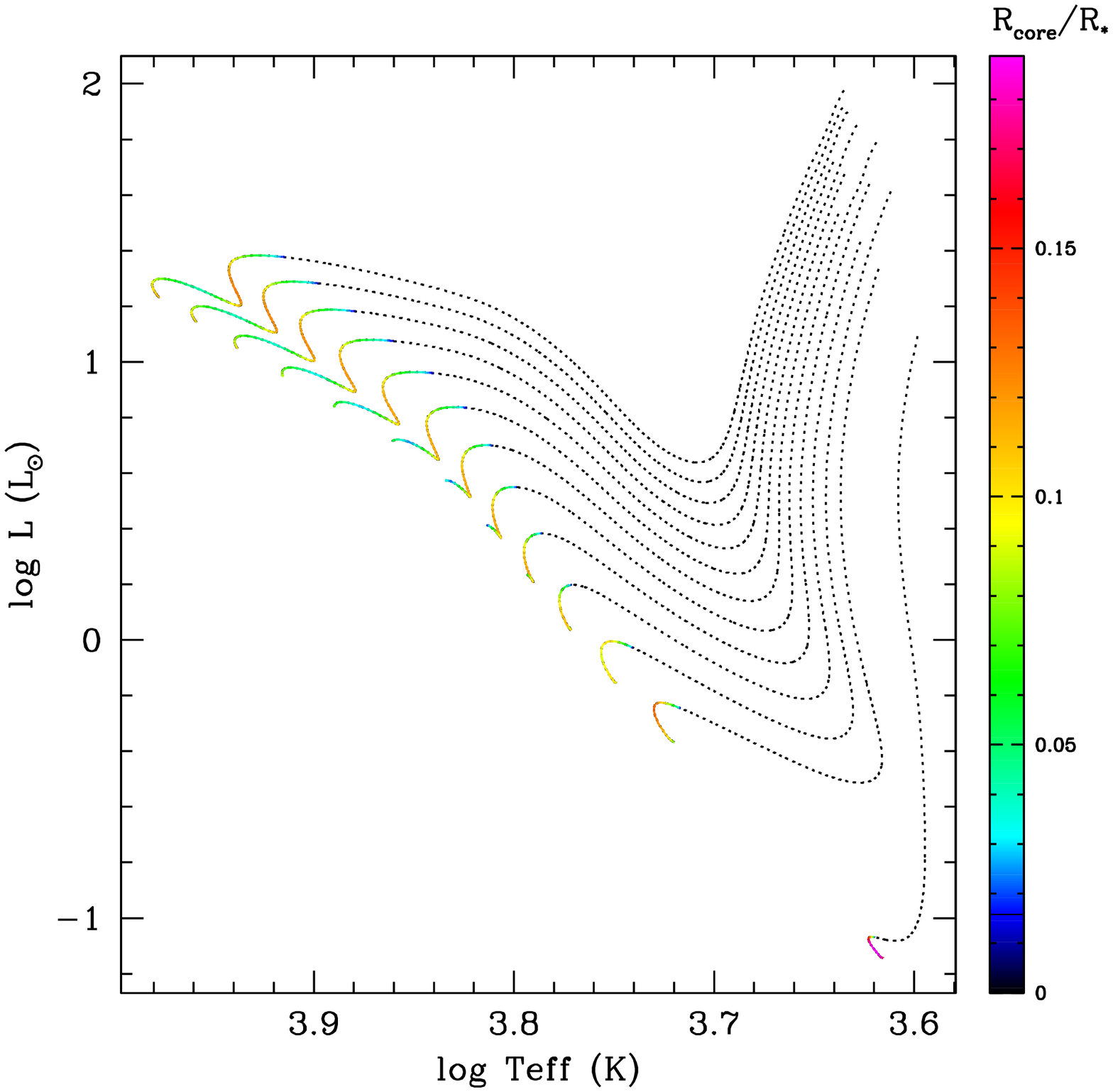}%
  \includegraphics[width=0.5\textwidth]{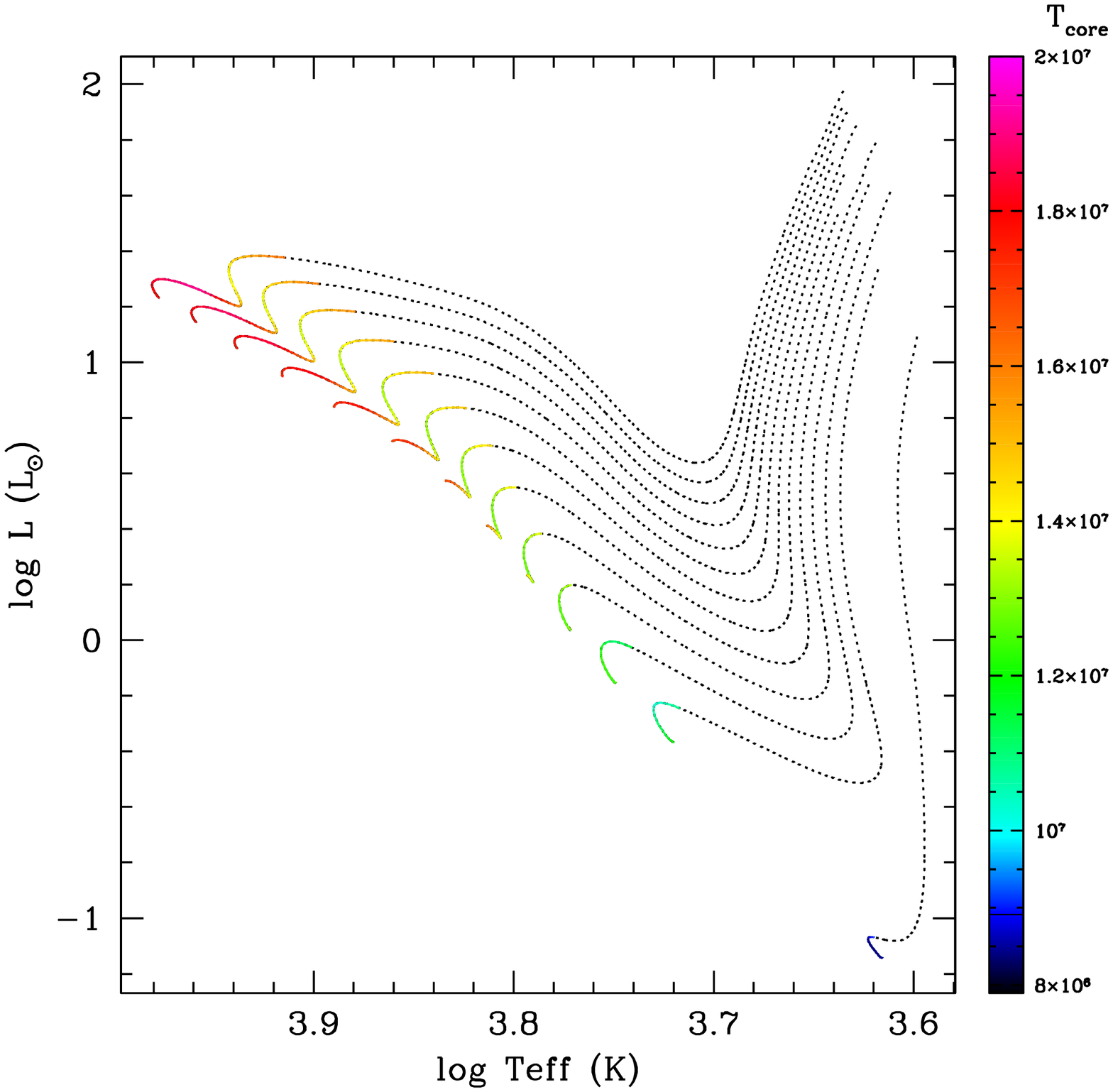}\\%
  \includegraphics[width=0.5\textwidth]{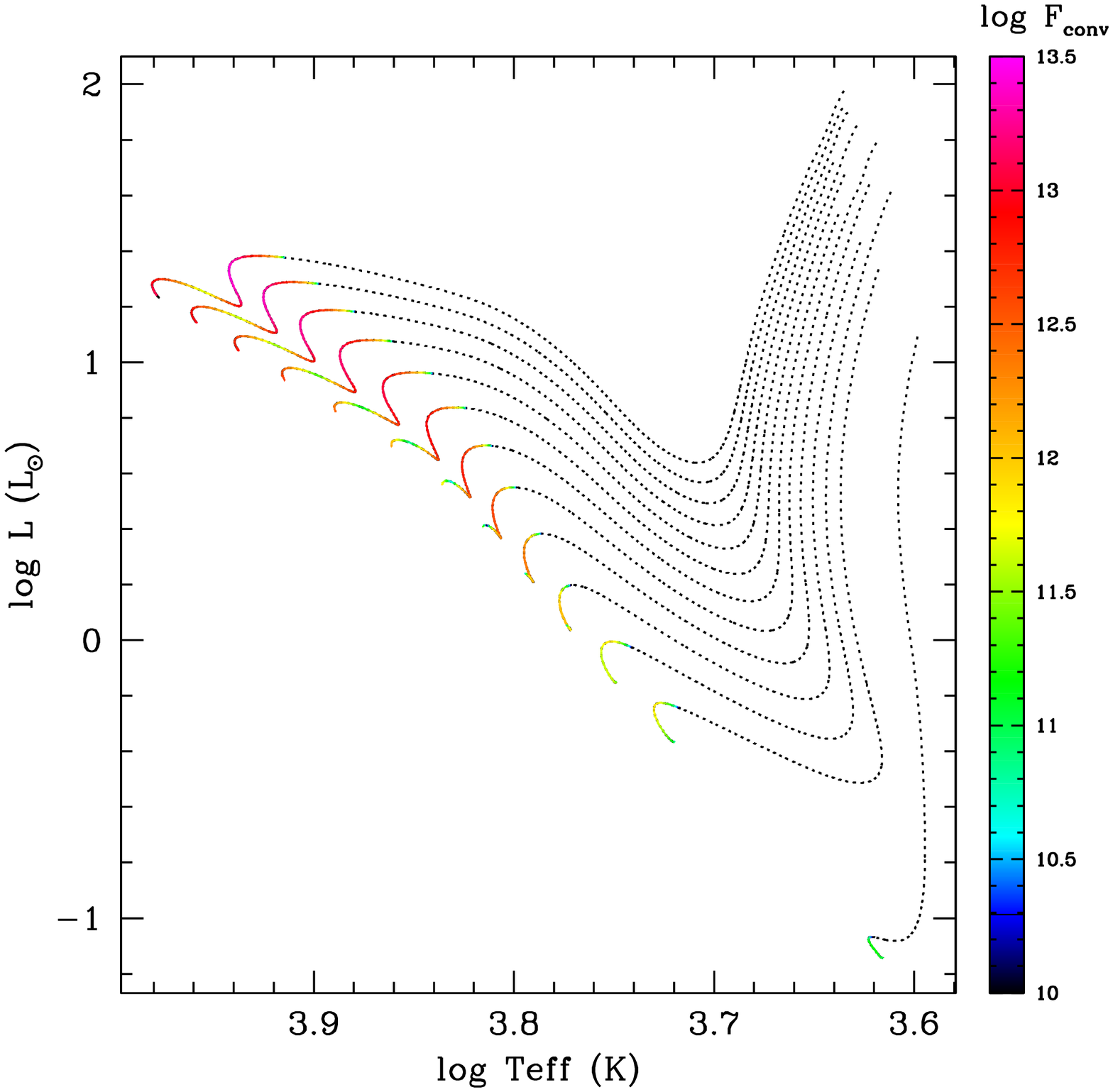}%
  \includegraphics[width=0.5\textwidth]{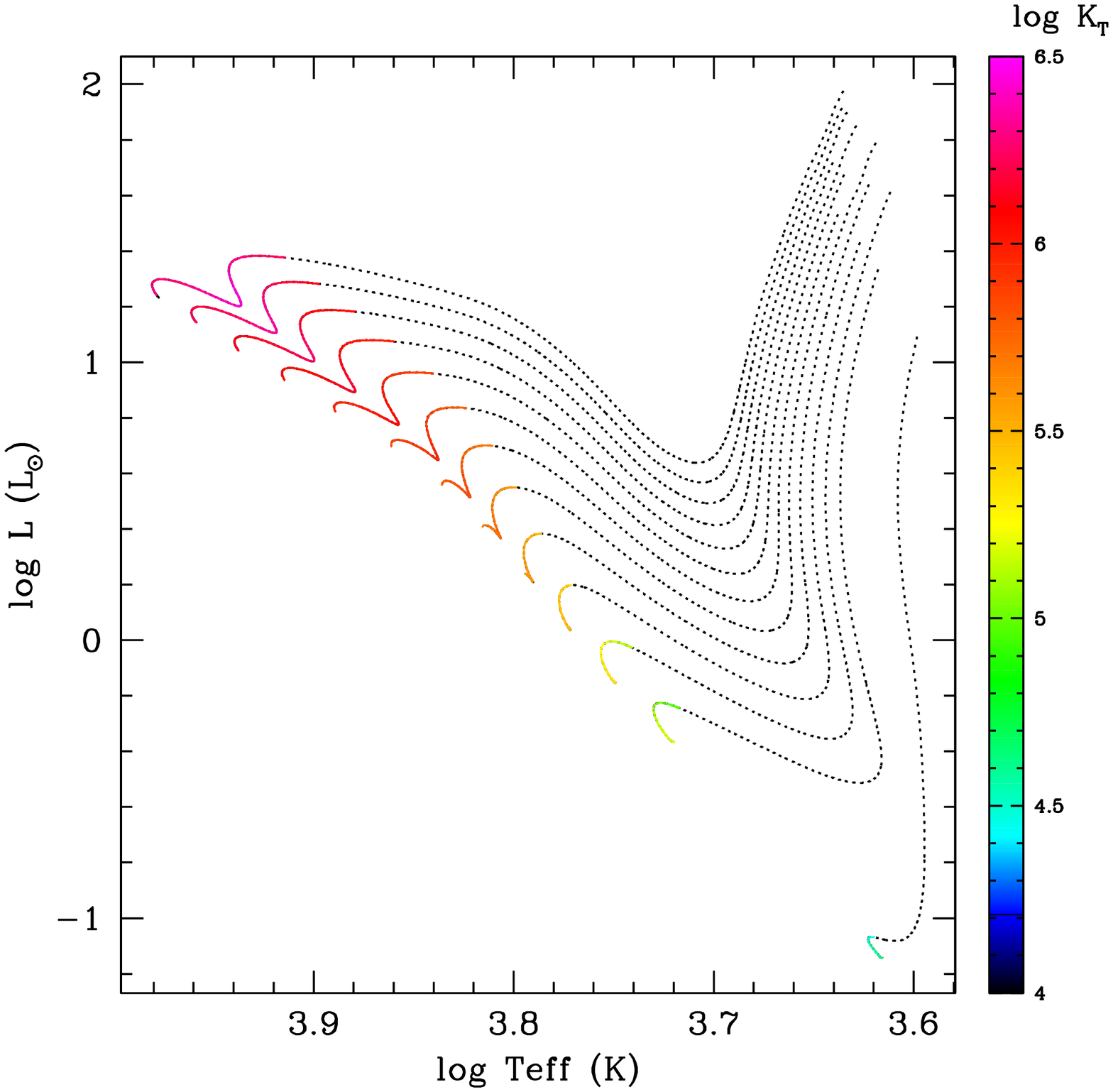}%
  \caption{Same as Fig.~\ref{fig:surfdetails}, but for the properties of the convective core. Here dotted parts on the tracks indicate the phase when the convective core is not yet present}
  \label{fig:coredetails}
\end{figure*}

\subsection{Initial internal and surface rotation}
\label{subsec:rotation}

We assume solid-body rotation while stars are fully convective (which corresponds to the dotted part of the tracks in Fig.~\ref{fig:surfdetails})  and we start computing the evolution of surface and internal rotation under the action of stellar contraction, meridional circulation, turbulence, and IGW when the radiative core appears, 
which happens at ages between $\sim$ 0.5 and 7.5~Myr for the mass range considered (see $\tau$(core) in Table~\ref{table:properties}), and at $\sim$2.5~Myr for the 1.0~M$_{\odot}$ model.
At that time, most or even all low-mass stars have already lost their disks as shown by observations in very young clusters \citep[e.g.][]{Haischetal01,Hartmann05,Hernandezetal08}. 
For all stellar masses, we choose the initial rotation velocity at the moment when the radiative zone appears to be equal to 5$\%$ of the critical velocity of the corresponding model 
(V$_{crit} = \sqrt{\frac{2}{3} \frac{G M} {R}}$; 
see Table~\ref{table:properties}). This corresponds approximately to the median of the observed distribution in young open clusters (see Fig.~\ref{fig:surfacerotation} and \S~\ref{section:globalproperties} for discussion). 
We assume that there is no more coupling between the star and a potential disk beyond that evolution point. The surface of the star is then free to spin up, and we do not apply any magnetic braking. 
The influence of the initial rotation velocity, of the disk lifetime that affects the moment when a PMS star starts spinning up,
as well as that of magnetic wind braking that may affect the rotation rate at the arrival on the ZAMS,
will be investigated in a forthcoming paper. 

\begin{table*}
  \caption{Properties of the different models computed without rotation (std), with rotation but without IGW (rot), and with rotation taking into account IGW (igw); for the 1.0~M$_{\odot}$ star a couple of models were computed taking into account the hydrostatic effects of rotation (rot+hydro and igw+hydro). 
   For each model we give: lifetime on the PMS, age at which the radiative core appears, initial rotation velocity, surface rotation rate and rotation period when the radiative core appears (taken at 5$\%$ of critical rotation velocity of the corresponding model), and surface rotation velocity, surface rotation rate, rotation period, surface lithium abundance, effective temperature, and luminosity at the arrival on the ZAMS}
      \centering
      \begin{tabular}{| c | c | c | c | c | c | c | c | c |c |c |c |c | }
    \hline
    Star  & & $\tau$(pms)   & $\tau$(core)  & v$_{surf}$  & $\Omega_{surf}$ / $\Omega_{\odot}$ & P$_\text{init}$ & v$_{surf}$   & $\Omega_{surf} / \Omega_{\odot}$ & P$_\text{ZAMS}$ & N(Li)   & Teff  & log L/L$_{\odot}$\\
    M$_{\odot}$ & & Myr & Myr & init, km.s$^{-1}$ & init & days & zams, km.s$^{-1}$ &  zams & days &zams & zams, K & zams \\
   \hline
0.6 & std      & 191 &7.47& --    &--    &-- &--    &--    & --    & -4.96 & 4118 & -1.14\\
    &rot      & 194  &7.47& 14.5  & 6.60 & 3.73 & 36.3  & 34.5 & 7.30  & -5.08 & 4127& -1.14\\
    &igw       & 217 &7.47& 14.5  & 6.60 & 3.73 & 37.3  & 35.6 & 7.12  & -5.40 & 4121& -1.14\\
\hline
0.9 & std      &  75    &3.00&--    &--    &-- &--    &--    & --    &     2.66  & 5246 &-0.366 \\
 & rot      &     76    &3.00&14.1& 4.16 & 5.99 &  48.5 & 30.7 & 0.83 &  2.64&   5246 & -0.366 \\
 & igw    &     89    &3.00&14.1& 4.16 &  5.99 & 55.7 & 35.2 & 0.72 &    2.64  & 5246 &-0.366 \\
\hline
1.0& std    & 57.5&2.43 &    --& --   &--    &-- &--    &     --     &    2.96  & 5608  &-0.152 \\
 & rot    & 53.9&2.43&    14.2 & 3.76& 6.84 & 53.1 & 29.8& 0.76 &     3.02  &
 5608 & -0.152 \\ 
 & rot+hydro& 59.4 &2.43& 14.2 & 3.76& 6.84 & 52.9 & 29.7 &  0.76  & 3.12 & 5589 & -0.152 \\
 & igw    & 59.5&2.43&    14.2 & 3.76& 6.84 & 63.1 & 35.5& 0.68 & 2.95 &  5608&  -0.152 \\
 & igw+hydro & 59.2   &2.43&  14.2 & 3.76 & 6.84 & 63.1 & 35.3   & 0.72   & 2.95
 &  5584 & -0.156 \\ 
\hline
1.2& std     &38.6&1.58&    --&    -- &-- &-- &--    &     --    &     3.17   &6217 & 0.235 \\
 & igw    & 43.5&1.58&    17.2& 2.99 & 8.03& 86.7  & 38.6 & 0.86&  3.17 &  6217 & 0.235 \\
\hline
1.4& std    & 26.5&1.09&    --&    -- &-- &-- &--    &     --    &     3.22  & 6828 &0.574 \\
 & igw    & 29.4&1.09&    13.6&  4.08  & 6.02 & 109  & 39.9 & 0.64&  3.22  & 6838 & 0.572 \\
\hline
1.6& std    & 19.6&0.81&    --& -- &--    &-- &--    &     --     &    3.23   &7757&  0.835 \\
 & rot    & 19.3&0.81&    13.0& 1.85 & 13.7& 100 & 39.2  & 0.71&  3.23  & 7757  &0.837 \\
 & igw    & 21.2&0.81&    13.0& 1.85 & 13.7& 120 & 42.2  &  0.57 & 3.23   &7783  &0.829 \\
\hline
1.8& std    & 15.0&0.62&    --&    -- &-- &-- &--    &     --    &     3.24&   8666 &1.04 \\
 & rot    & 15.0&0.62&    12.4& 1.43& 16.9& 112 & 37.6& 0.68&  3.24  & 8666  &1.04 \\
 & igw    & 16.1&0.62&    12.4& 1.43 & 16.9& 128    & 43.9& 0.57   &     3.24   &8676&  1.04 \\
\hline
2.0& std    & 11.8&0.48&    --& --     &-- &-- &--    &     --     &    3.24  & 9479 & 1.23 \\
 & rot     &11.7&0.48&    20& 3.32    & 7.62& 124  & 40.7   & 0.62&  3.24  & 9473 & 1.23 \\
 & igw    & 12.1&0.48&    20& 3.32    & 7.62& 139  & 46.0   & 0.54&  3.24  & 9492 & 1.23 \\
\hline
    \end{tabular}
\label{table:properties}
\end{table*} 

\section{IGW generation along the PMS evolution for all grid models}
\label{sec:IGWexcitation}

The internal structure strongly changes as low-mass stars evolve along the PMS. 
This implies strong variations in the quantities that are relevant to IGW generation and momentum transport, as depicted in Figs.~\ref{fig:surfdetails} for the properties of the convective envelope and \ref{fig:coredetails} for the core. 
All quantities are given in cgs units.

Stars are first fully convective and a radiative core appears along the Hayashi track as they contract and heat (Fig.~\ref{fig:surfdetails}). 
The thickness of the convective envelope decreases, and the temperature at its base increases as the stars move towards higher effective temperatures.
Due to central CNO-burning ignition on the final approach towards the ZAMS a convective core develops (Fig.\ref{fig:coredetails}).

\begin{figure*}
  \includegraphics[width=0.5\textwidth]{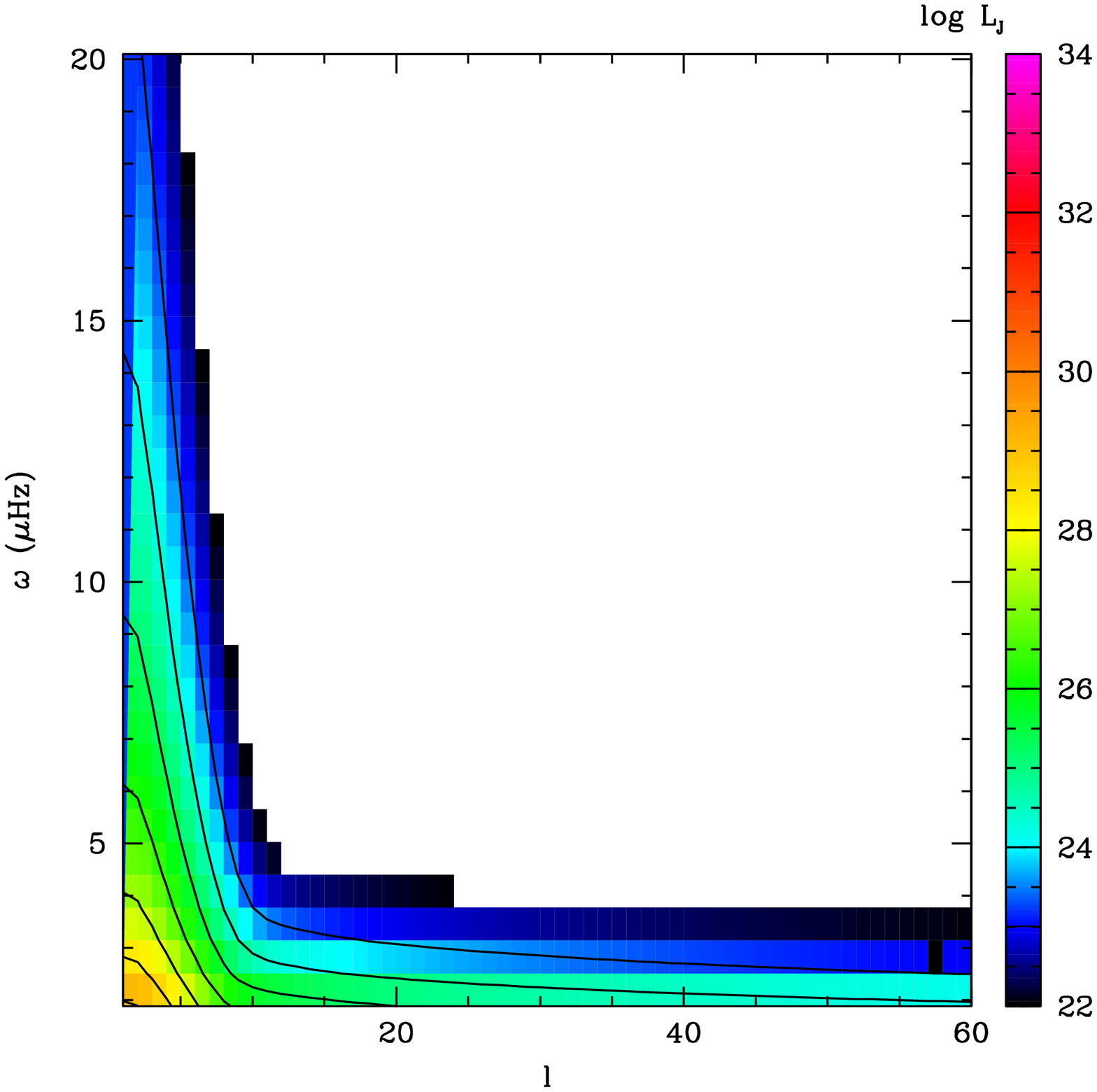}
  \includegraphics[width=0.5\textwidth]{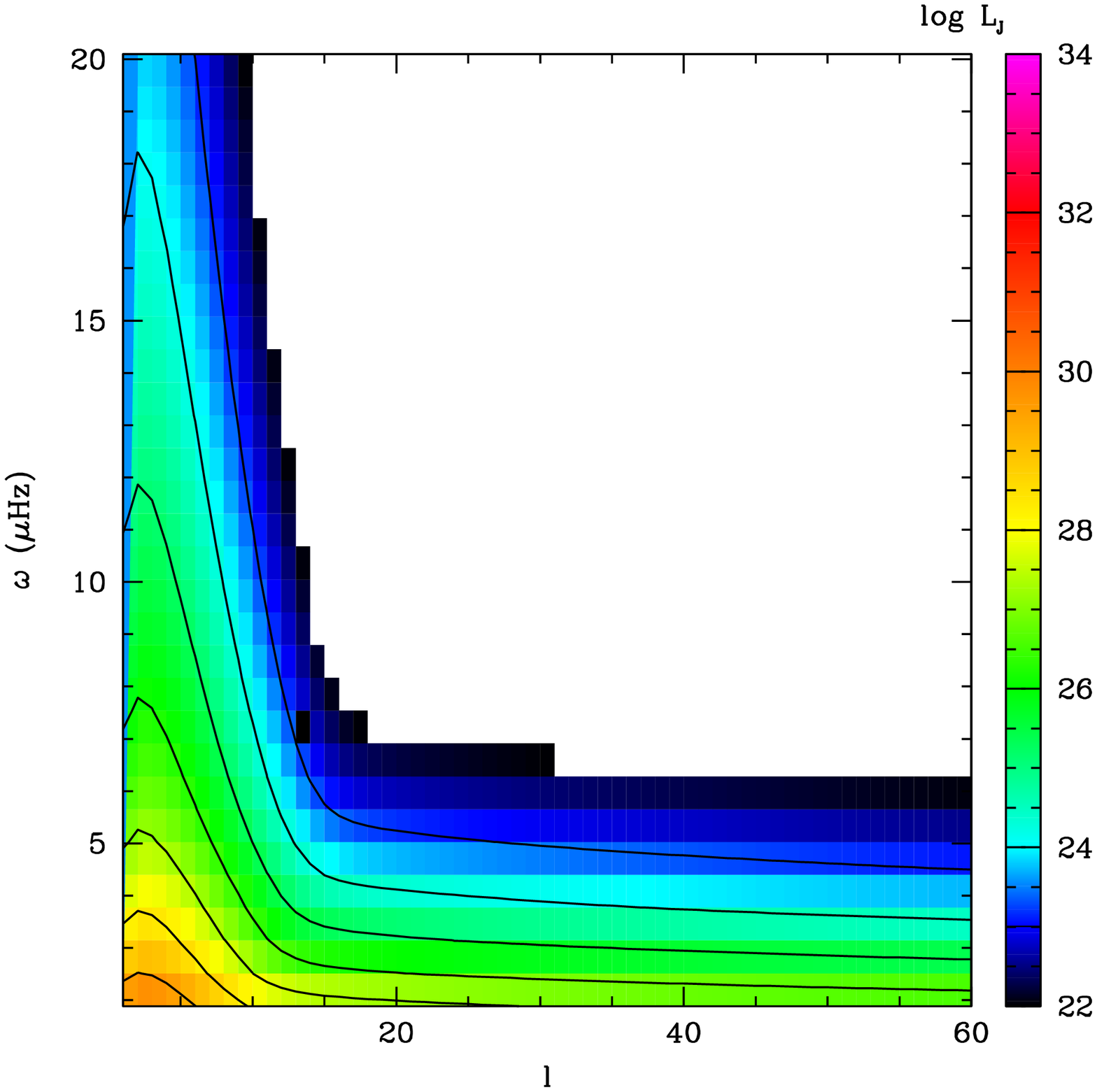}
  \includegraphics[width=0.5\textwidth]{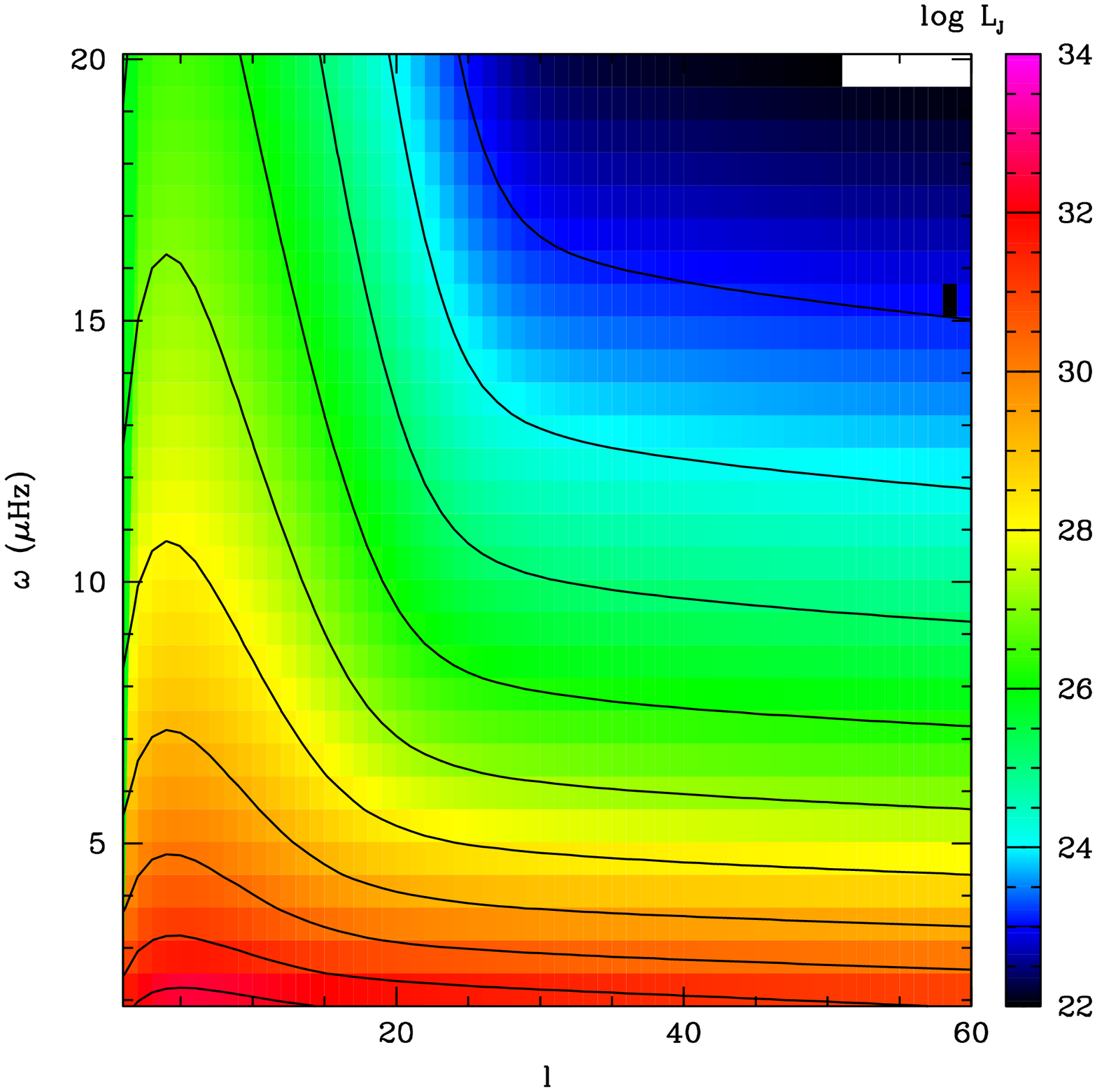}
  \includegraphics[width=0.5\textwidth]{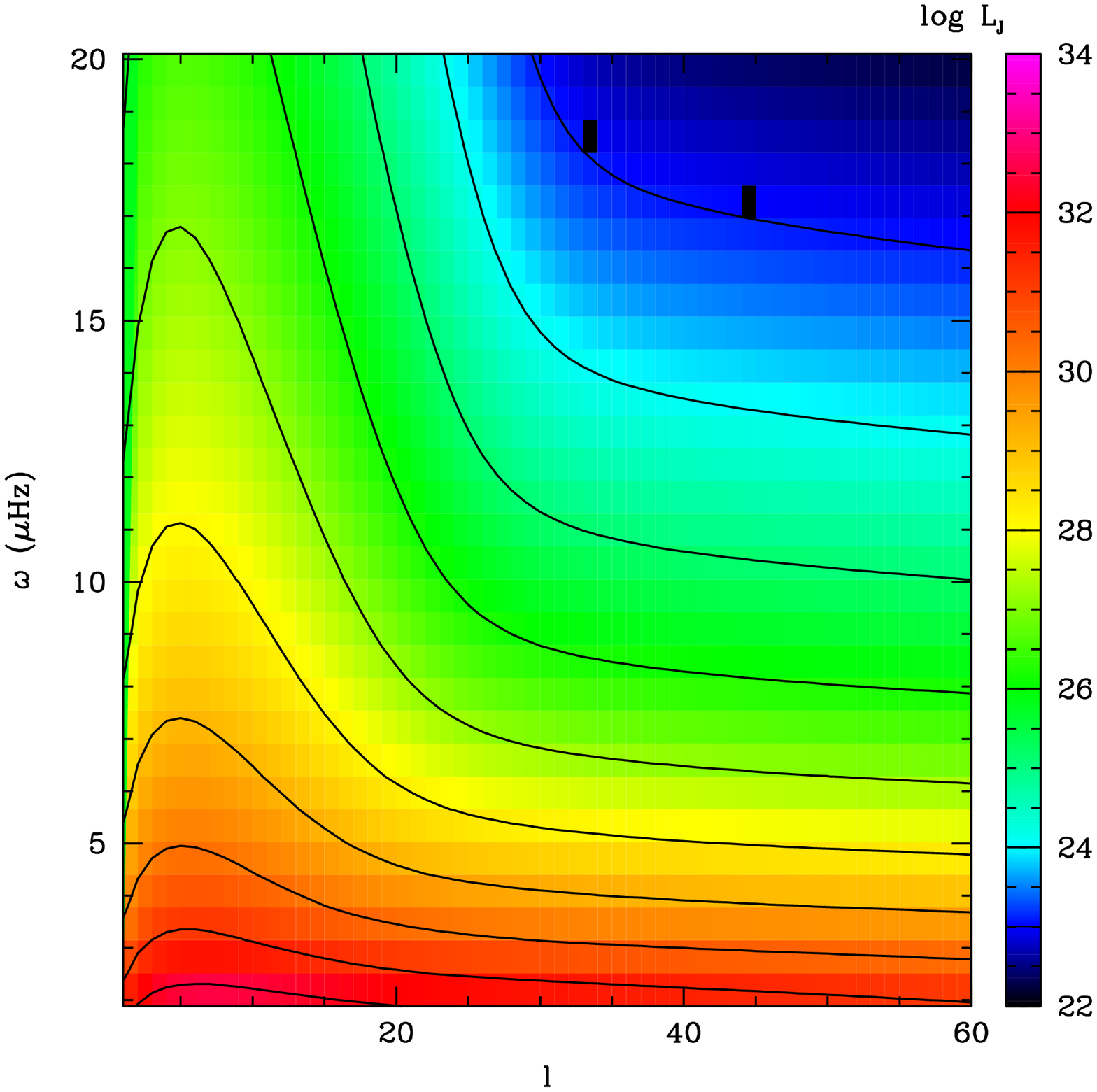}
  \caption{Angular momentum luminosity at the base of the convective envelope of IGW generated by Reynolds-stress in the external convective layers  
  as a function of emission frequency $\omega$ and degree $\ell$. The color axis is in log and white areas correspond to log${\cal L}_{J, surf} <22$. The plots are shown for the 1~M$_{\odot}$ model at four ages along the PMS (5.8, 14, 35, and 55~Myr from top left to bottom right; the corresponding values of T$_{eff}$ are 4277, 4357, 5560, and 5612~K). 
  }
  \label{fig:igwexcitspectrum1Msun}
\end{figure*}

\begin{figure*}
  \includegraphics[width=0.5\textwidth]{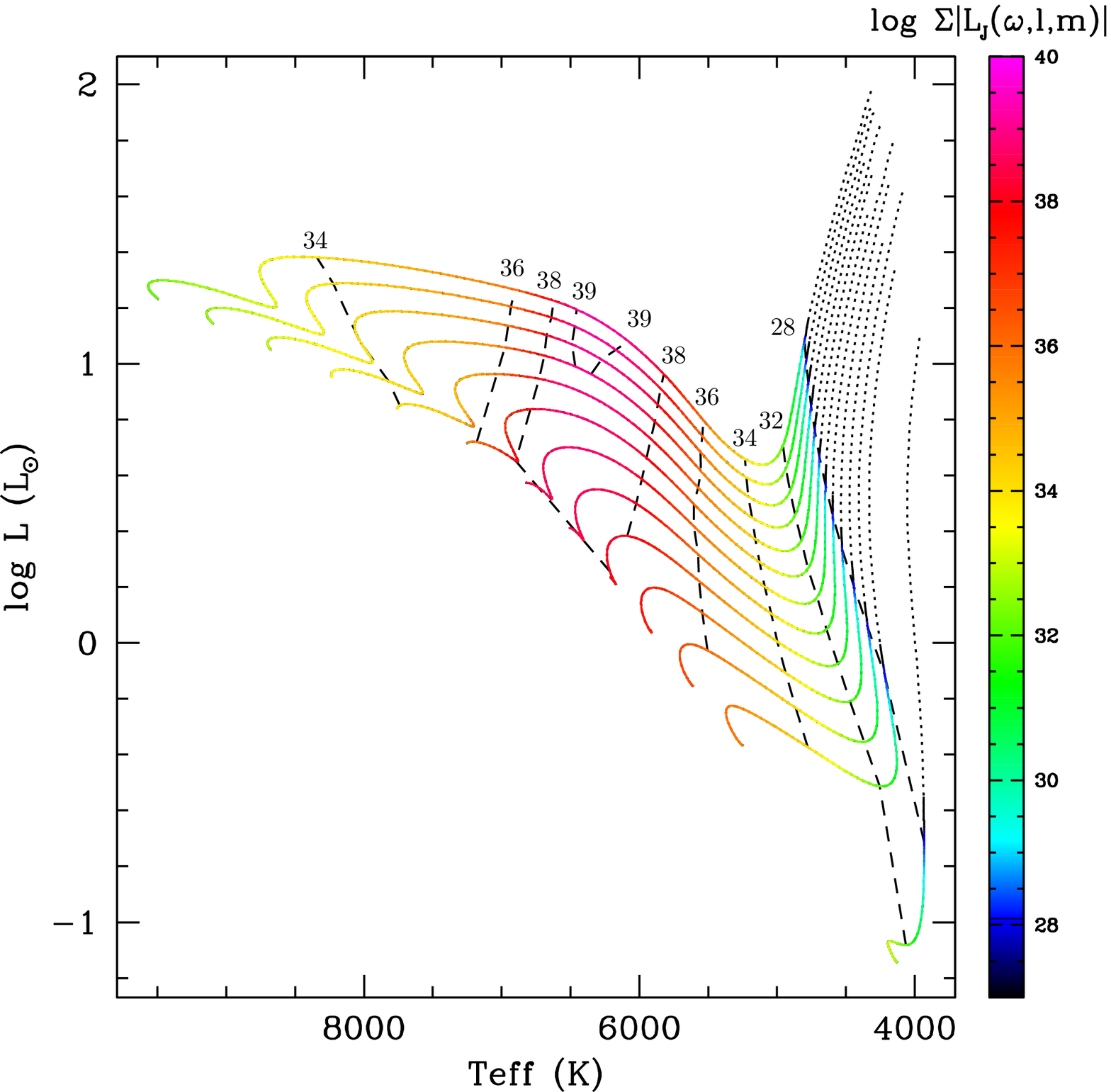}%
  \includegraphics[width=0.5\textwidth]{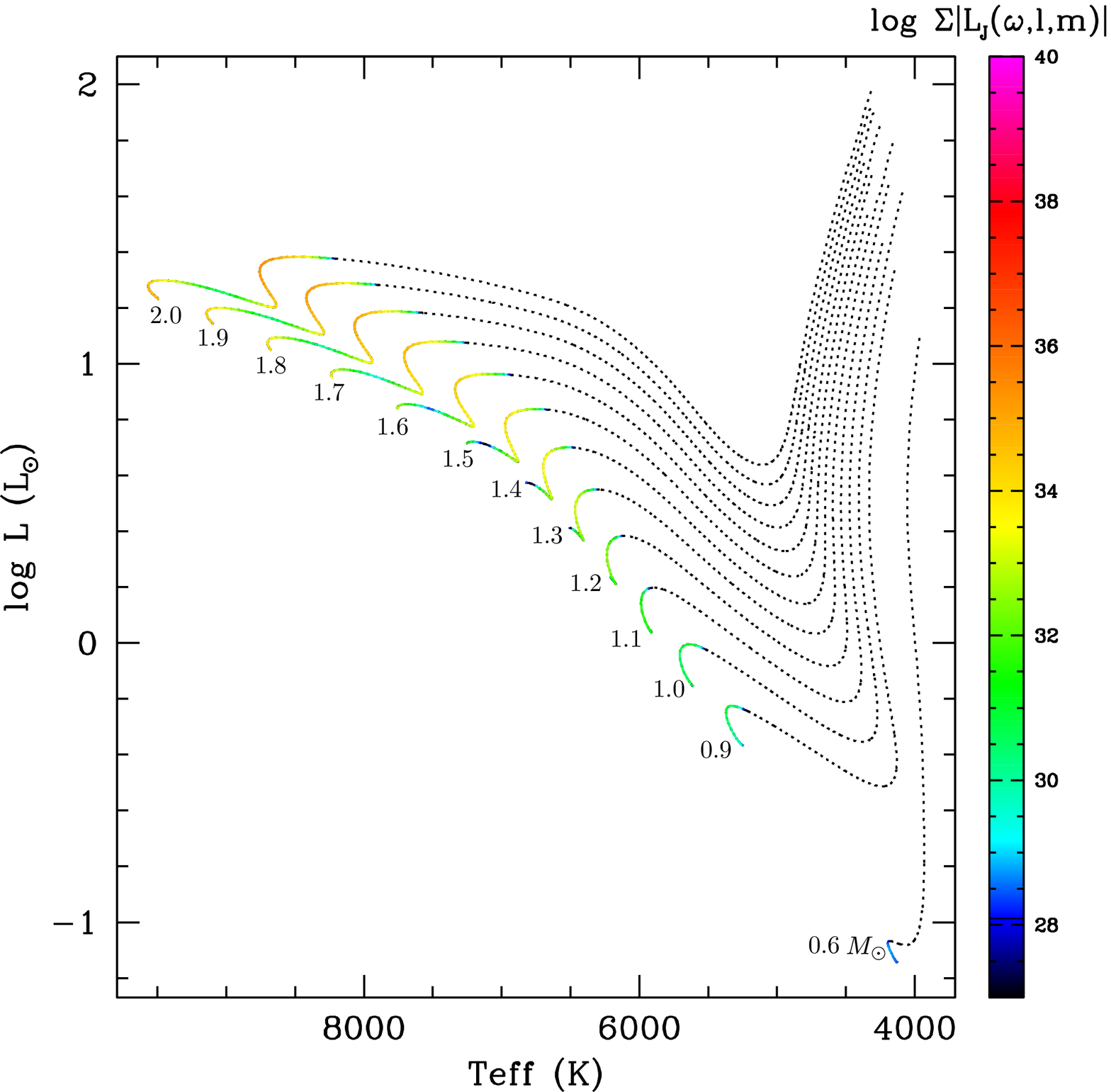}%
  \caption{Same as Fig.~\ref{fig:surfdetails}, but with colors indicating the total momentum
    flux carried by IGW generated by the convective envelope (left) and
    the convective core (right). Dotted lines indicate the region where the
    stars are fully convective (left) or have no convective core (right) so
    that no IGW can be generated. In the left panel the vertical dashed
    lines connect models where the excitation has the same value: $\log \left(\Sigma
    |\mathcal F_J(\omega,l,m)|\right) = 28$, 30, 32, 34, 36, 38.}
  \label{fig:igwexcit}
\end{figure*}

One can follow the evolution along the tracks of the maximum convective flux ($F_c = C_p \rho v_c \Delta T$) inside the external and central convective regions, 
which directly affects the energy flux associated to a given frequency (see Eq.\ref{gold}).
Wave excitation is stronger when the convective length scale ($\ell_c = 2 \pi r_{cz} / \alpha H_p$) is larger, but decreases when the turnover timescale ($\tau_c = \alpha_{MLT} H_p / v_c$)
becomes too large. 
The combination of these two factors induces large differences in the overall efficiency of wave generation as the internal structure evolves.
This is well illustrated in Fig.~\ref{fig:igwexcitspectrum1Msun} that shows the luminosity spectrum of IGW generated by the external convection zone in the 1~M$_{\odot}$ model at four ages on the PMS.
One sees clearly that wave-induced transport is dominated by low-frequency waves (i.e., $< 3.5~\mu$Hz). 
High degree waves at  low frequencies do not contribute much to the transport of angular
  momentum even though their excitation flux is important : indeed they are essentially
  damped near the convective envelope edge.
In Fig.~\ref{fig:igwexcit} colors along the tracks indicate the net momentum luminosity ${\cal L}_J$ (see Eq.~\ref{eq:angmomluminosity})
of IGWs generated by the external and internal convective regions. 

In the case of external convection, the net momentum luminosity ${\cal L}_{J, surf}$ rapidly increases as the excitation of IGW strengthens up when 
stars evolve towards higher effective temperature, and reaches maximum values as high as 10$^{39}$~g.cm$^2$s$^2$ around $T_\text{eff} \sim 6200$~K. 
Stars with initial masses lower than 1.3 \Ms{} never reach this effective temperature and the corresponding ${\cal L}_{J, surf}$ remains always
below this maximum and shows only a monotonic increase along the PMS.
On the other hand in the more massive models the convective envelope keeps shrinking in size and ${\cal L}_{J, surf}$ decreases when T$_{eff}$  increases above 6200~K.
This behavior confirms TC08 findings for intermediate-mass PMS stars, and is very similar to the ${\cal L}_J$ plateau we found for Pop I and Pop II main sequence stars \citep{TalonCharbonnel2003,TalonCharbonnel04},
 which share very similar convective properties with PMS stars in the same T$_{eff}$ range.

\begin{figure}
  \includegraphics[width=0.5\textwidth]{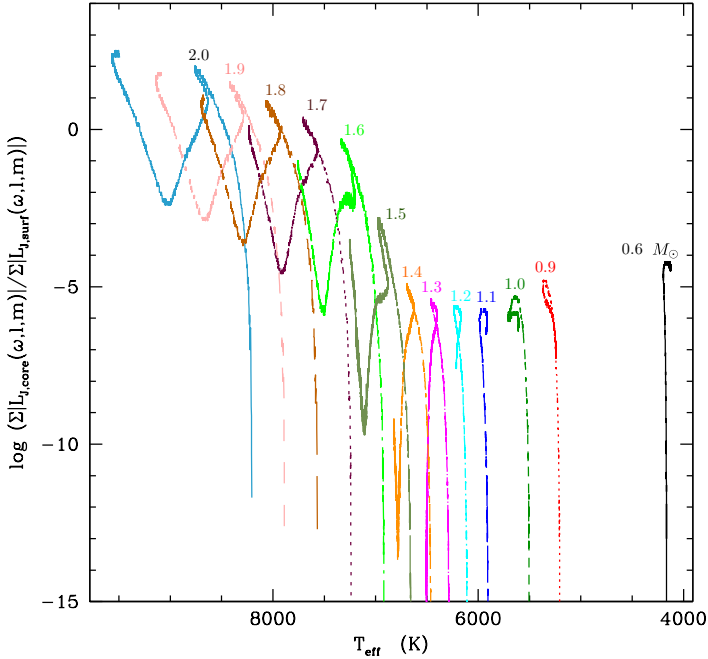}
  \caption{Ratio of the total momentum luminosity carried by IGW generated
    by the convective core and the convective envelope for the PMS models of various masses
  }
  \label{fig:excitrenv}
\end{figure}

IGW are also emitted from the convective core at the end of the PMS. 
The more massive the star, the more the convective core expands, and the stronger the corresponding wave excitation. 
We note however from Fig.~\ref{fig:igwexcit} that wave excitation by the convective core (when present) is generally much less efficient than that of the convective envelope. 
The ratio between ${\cal L}_{J,core}$ and ${\cal L}_{J,surf}$ is shown in Fig.~\ref{fig:excitrenv} as a function of T$_{eff}$  for the various models.
For stars with masses below 1.4 \Ms, ${\cal L}_{J,core}$ is always $\sim$ 5-6 order of magnitude lower than ${\cal L}_{J,surf}$. 
These two quantities reach similar orders of magnitude only very close from the ZAMS for stars more massive than 1.6 \Ms{}. 
Therefore and as we shall see below, the impact of IGW on the internal rotation profile along the PMS will be dominated by the waves emitted by the convective envelope.

\section{Evolution of the internal rotation profile} 
\label{evolutionrotprofile}

\subsection{The case of the 1~M$_{\odot}$ star}
\label{subsection:1MZsun}

\begin{figure*}
	\centering
 \includegraphics[width=0.45\textwidth]{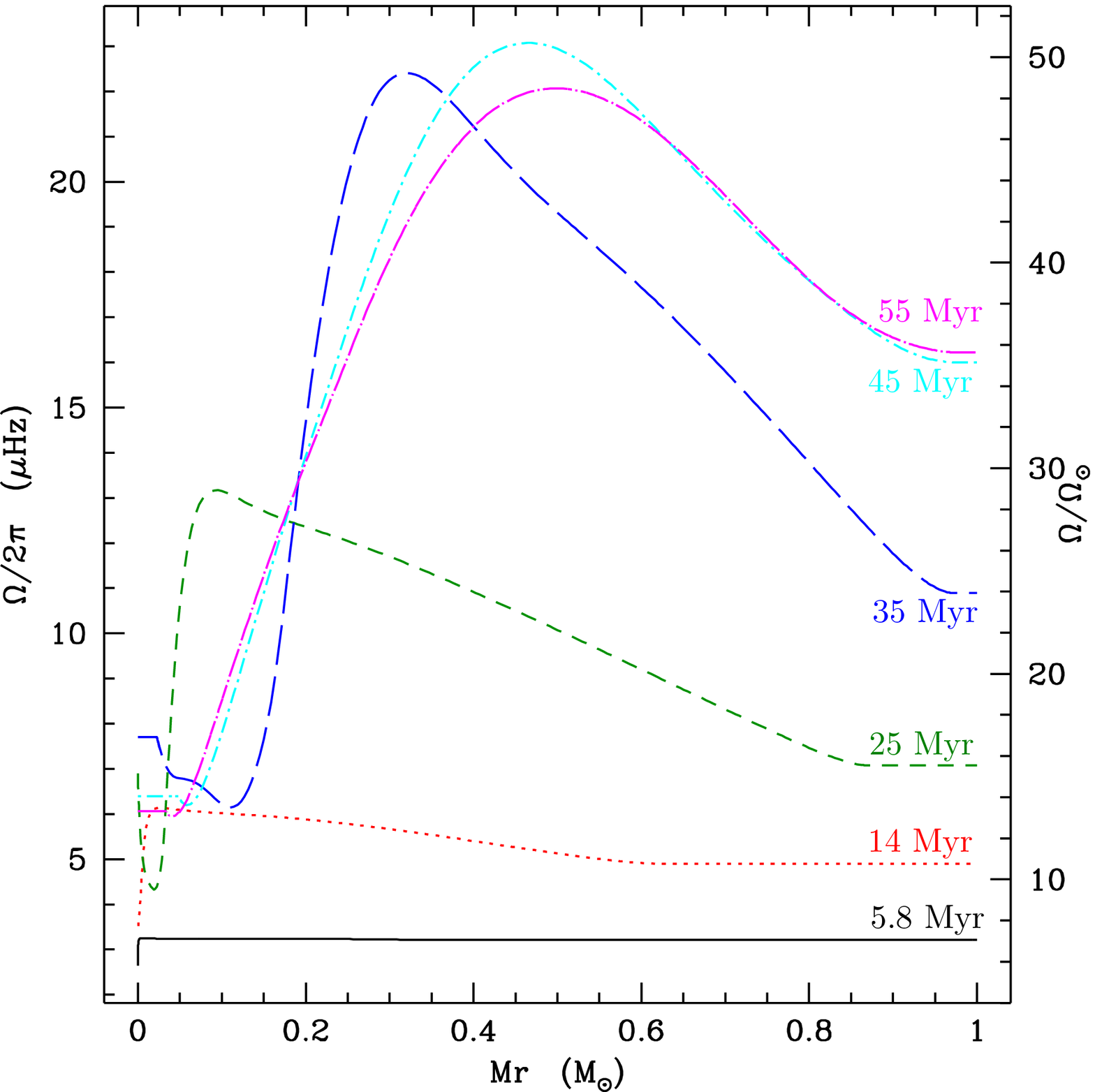}
  \includegraphics[width=0.45\textwidth]{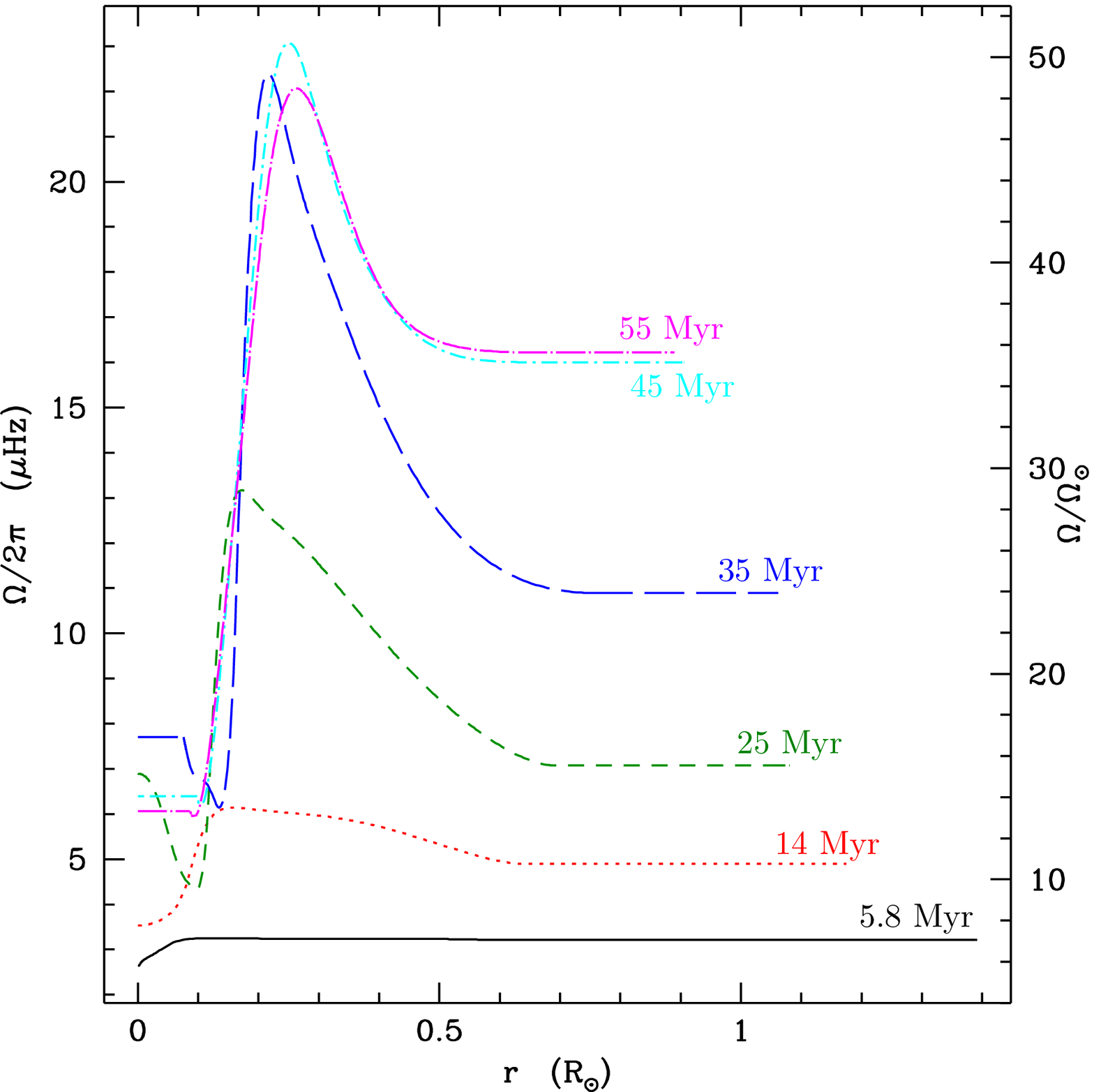}\\
    \includegraphics[width=0.45\textwidth]{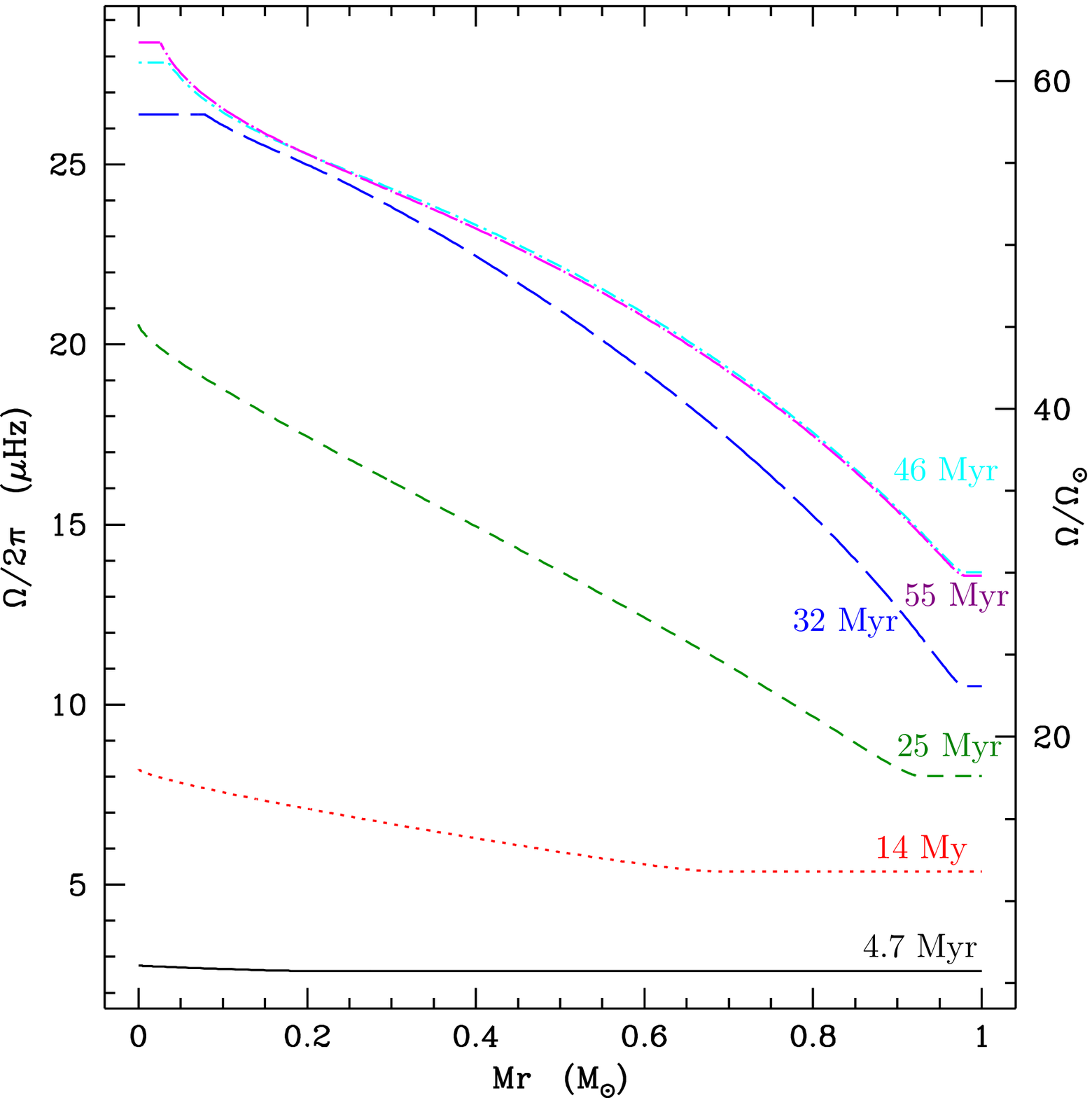}
  \includegraphics[width=0.45\textwidth]{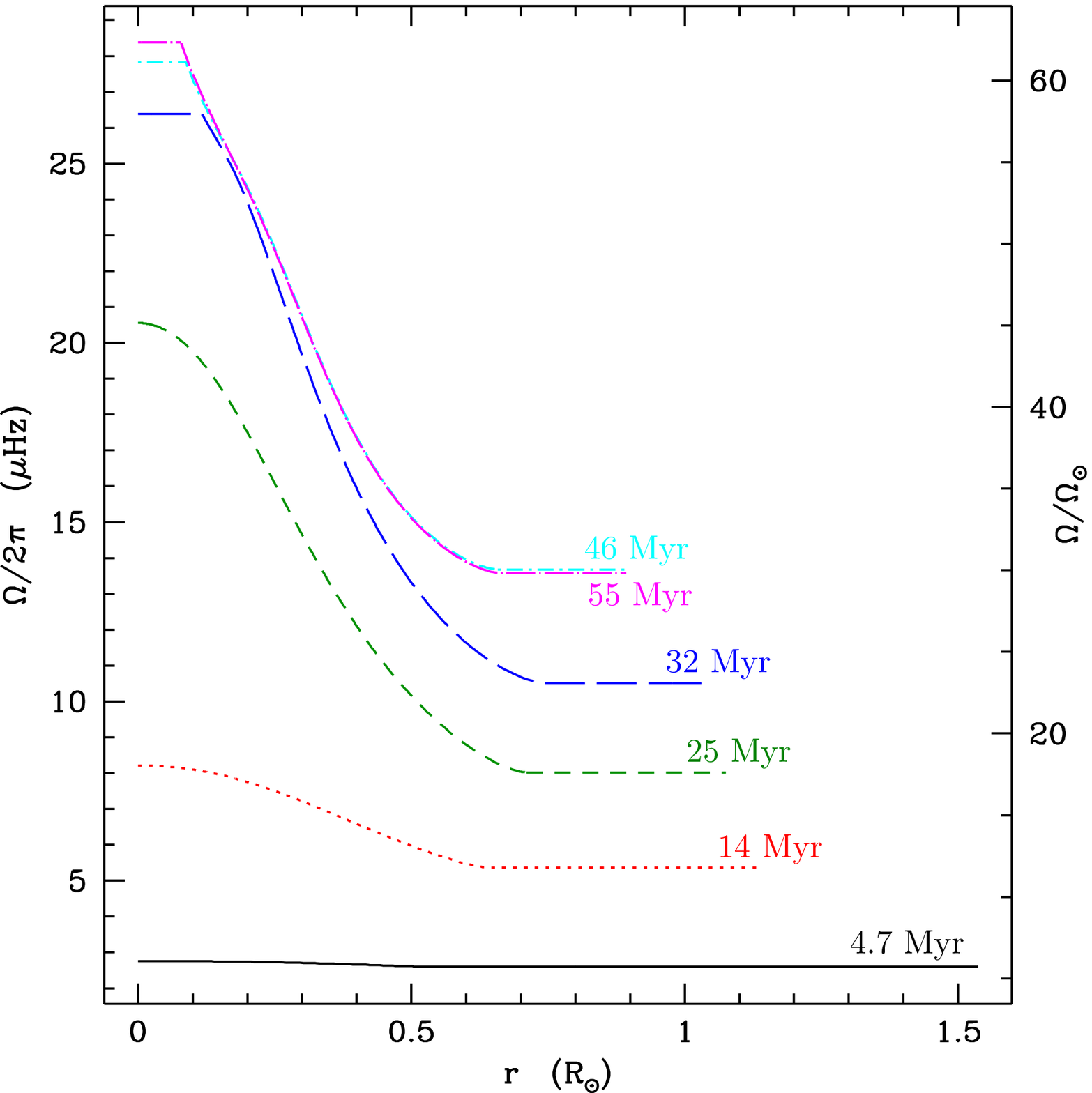}
    \caption{Evolution along the PMS of the rotation profile in the 1~M$_{\odot}$ models computed with and without IGW (top and bottom respectively) as a function of relative mass fraction and radius in solar units (left and right respectively). 
        The curves are labeled according to age. On each plot the left and right scales give $\Omega$ in $\mu$Hz and in solar units respectively}
   \label{fig:omegaHzevol1p0}
\end{figure*}

\begin{figure*}
\includegraphics[angle=0,width=8.5cm]{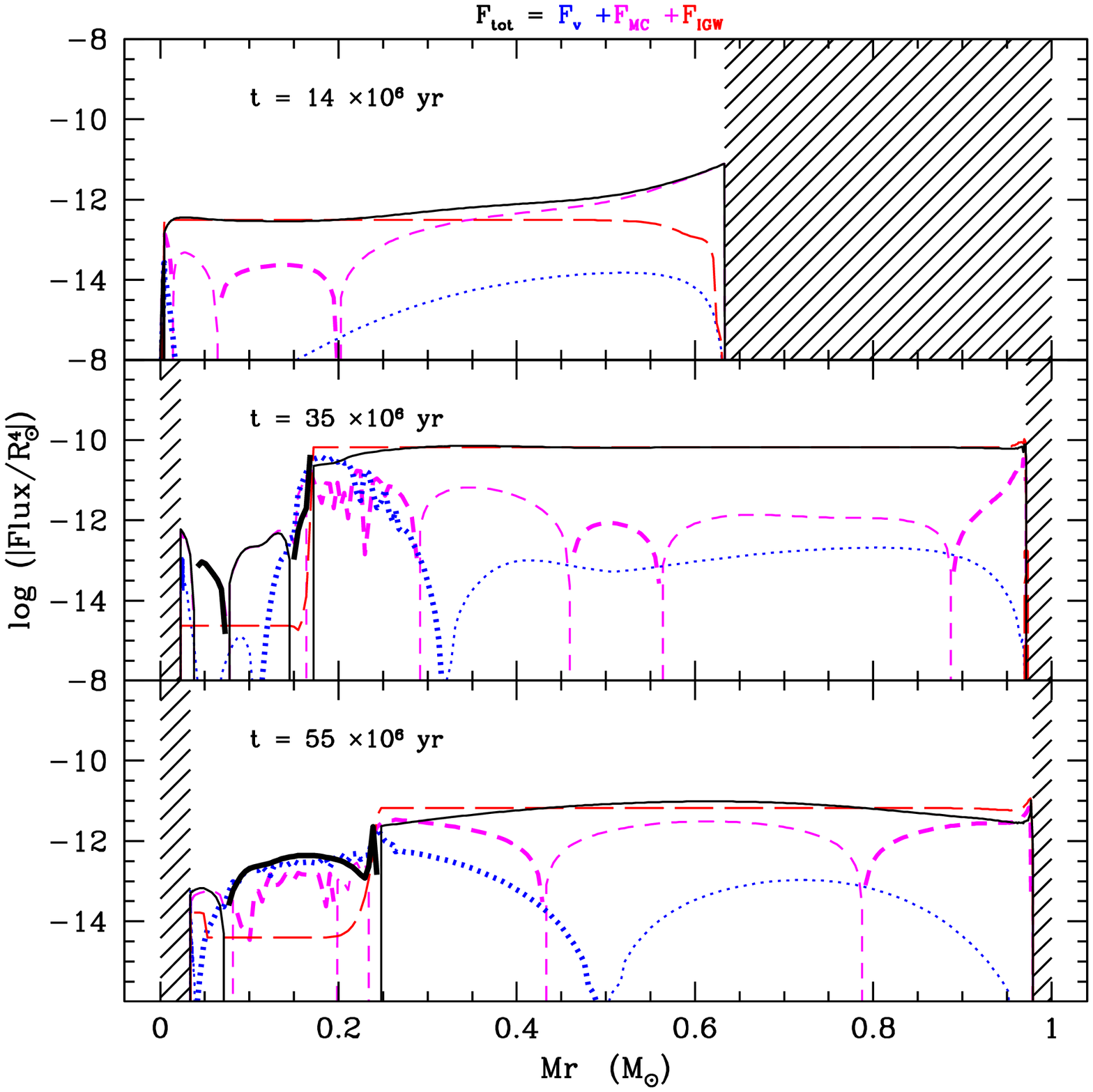} 
\includegraphics[angle=0,width=8.5cm]{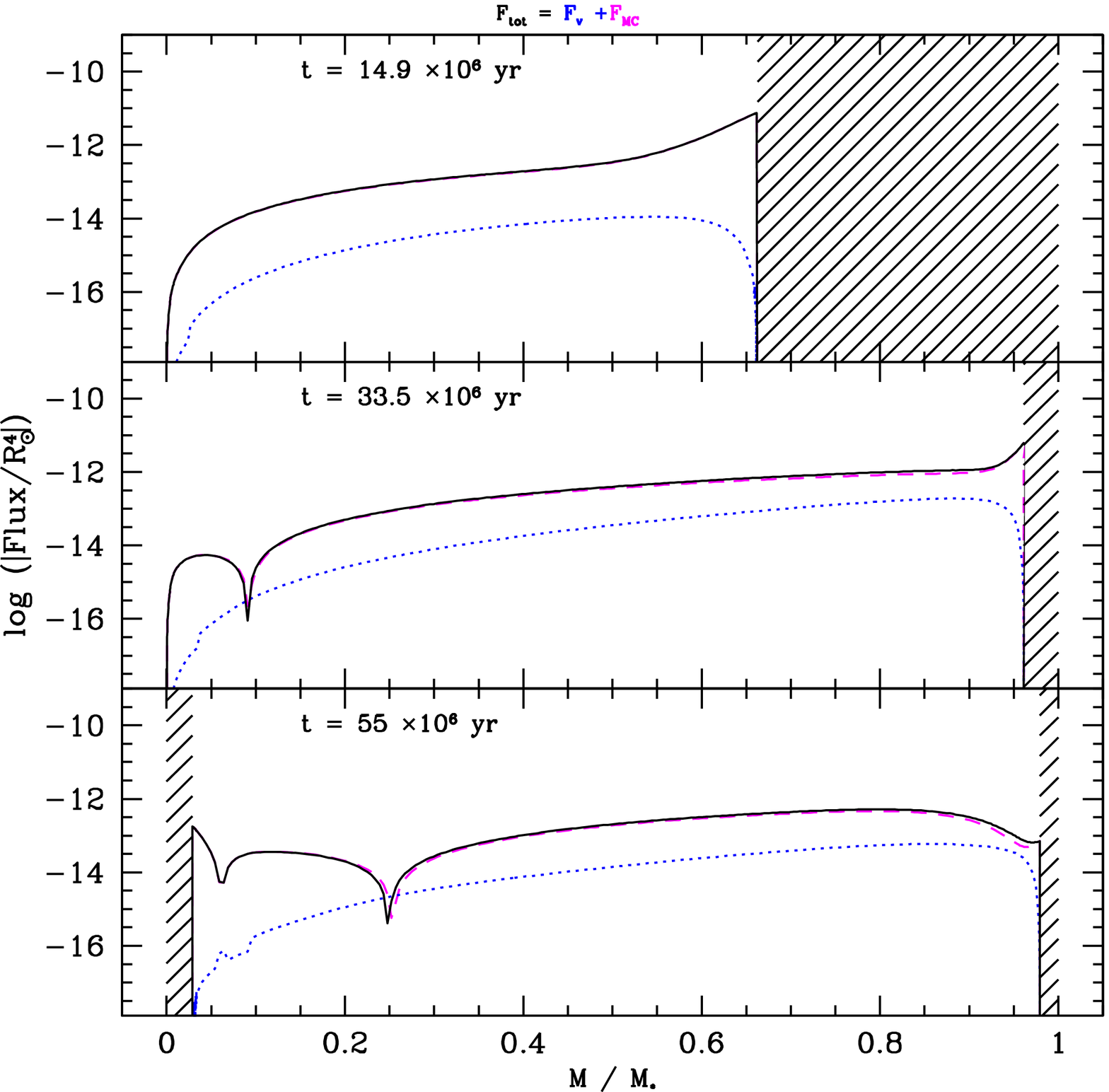}
\caption{Decomposition of the total flux of angular momentum (solid black) into meridional circulation (long-dashed magenta), shear turbulence (dotted blue), and IGW (short-dashed red) in the 1~M$_{\odot}$ models computed with and without  IGW (left and right panels respectively).  Bold lines indicate negative values for the fluxes $F_{\rm MC}\left(r\right)$, $F_{\rm S}\left(r\right)$, or $F_{\rm IGW}\left(r\right)$, when angular momentum is transported towards the central regions by the corresponding mechanism; in the case of meridional circulation and of shear turbulence this corresponds respectively to clockwise currents ($U_2 > 0$) and to a positive $\Omega$ gradient. The profiles are shown at three different ages along the PMS. Shaded areas correspond to convective regions}
\label{flux}
\end{figure*}

\begin{figure*}
\includegraphics[width=\textwidth]{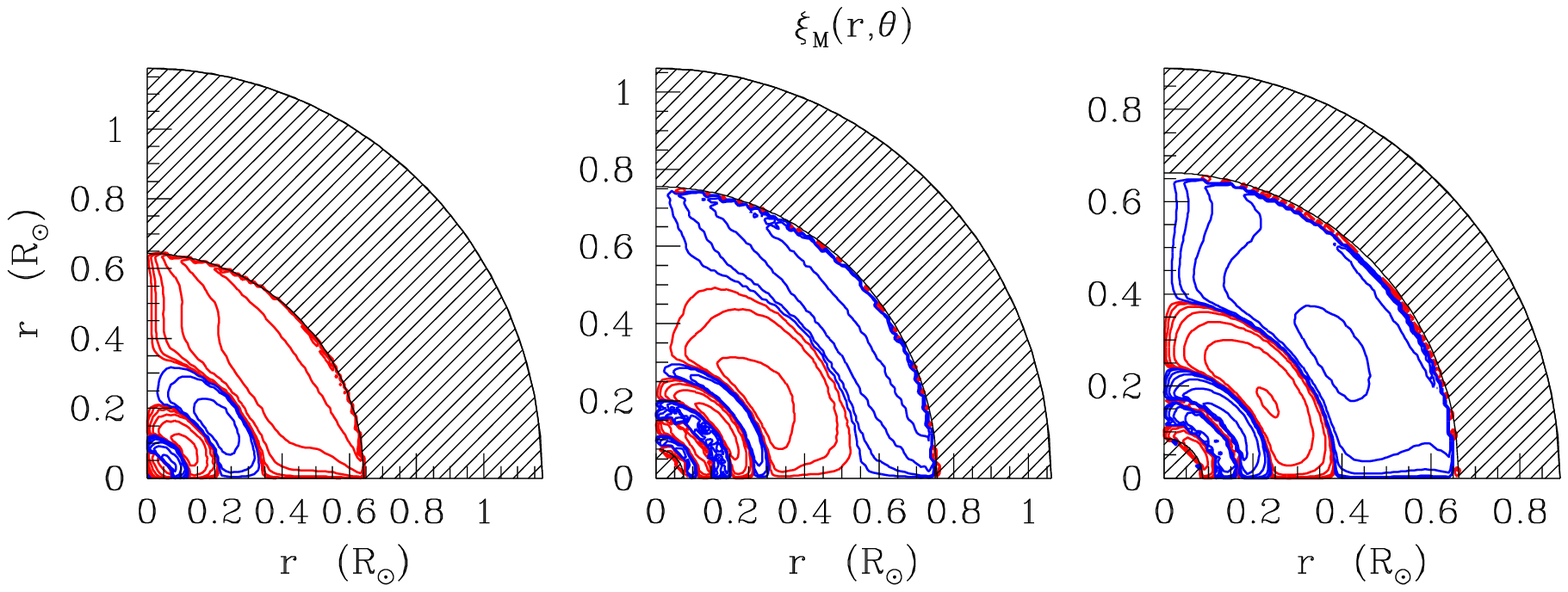}
\includegraphics[width=\textwidth]{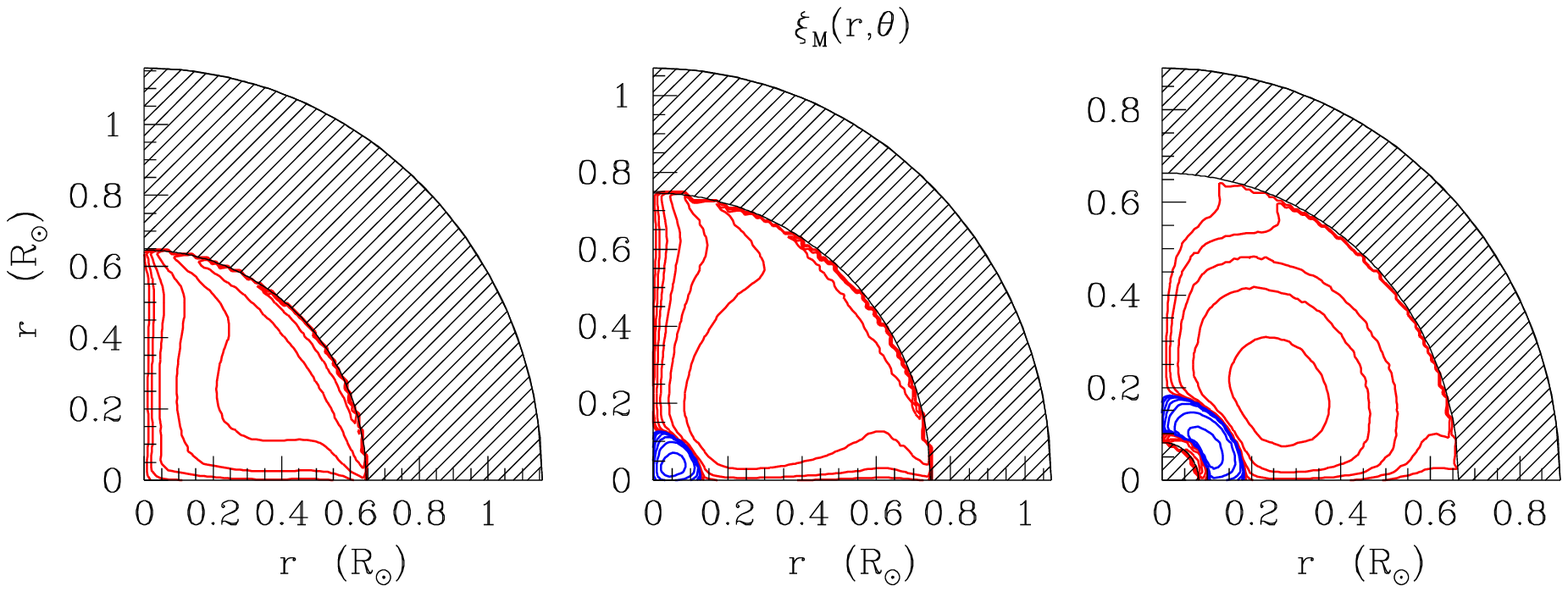}
\caption{Meridional circulation currents in  the 1~M$_{\odot}$ models computed with and without IGW (top and bottom respectively) at three different evolution ages along the PMS (14, 35, and 55Myrs from left to right)). Blue and red lines indicate clockwise ($U_2 > 0$) and counterclockwise ($U_2 < 0$) circulation respectively. Hatched areas correspond to convective regions}
\label{fig:U2}
\end{figure*}

Figure~\ref{fig:omegaHzevol1p0} depicts the evolution along the PMS of the rotation profile inside the 1~M$_{\odot}$ star for two cases: when angular momentum transport is operated solely by meridional circulation and shear turbulence (bottom panels), and when angular momentum deposition by internal gravity waves is taken into account in addition to the hydrodynamic processes (top panels); the rotation profile is shown at different ages as a function of both relative mass fraction and reduced radius (left and right panels respectively). 
The decomposition of the total flux of angular momentum into the various components driven by meridional circulation, shear turbulence, and IGW (when accounted for; see Eqs.~\ref{eq:Fmc}, \ref{eq:FS}, and \ref{eq:FIGW} respectively) 
is shown in Fig.~\ref{flux} at three ages along the PMS. 
Meridional circulation currents are shown at the same ages in Fig.\ref{fig:U2}; clockwise currents (matter flowing from the equator to the pole and resulting in deposition of angular momentum inwards) and counterclockwise ones (carrying angular momentum outwards) are drawn in blue and red respectively. 

\subsubsection{Transport of angular momentum by meridional circulation and shear turbulence only}
\label{subsubsect:mershear}
When only meridional circulation and shear turbulence are accounted for, 
differential rotation rapidly develops inside the radiative region as 
the surface rotation velocity increases due to stellar contraction (Fig.~\ref{fig:omegaHzevol1p0}, bottom plots). 
This behavior as well as the rotation profile at the arrival on the ZAMS are similar to the results of \citet{Eggenbergeretal12} for their rotating 1~M$_{\odot}$ model computed with similar assumptions.

As can be seen in Fig.\ref{flux} (right plots)  for this model without IGW, the transport of angular momentum is dominated by 
meridional circulation all along the PMS, while the contribution of shear turbulence is negligible (the flux of angular momentum by turbulence $F_V$ is indeed $\sim$ 2 orders of magnitude lower than the flux driven by meridional circulation $F_{MC}$). 
The number of circulation loops evolves with time (Fig.\ref{fig:U2}, bottom plots;  see also Fig.~\ref{flux}): In the early stages (14.9~Myrs, left panel),  the circulation consists of a single counterclockwise current that transports matter inward along the rotational axis and outward in the equatorial plane; later on (33.5~Myrs, middle panel) a clockwise loop appears in the central regions; finally an additional counterclockwise loop shows up when the convective core develops (55~Myrs, right panel).

\subsubsection{Impact of internal gravity waves}
\label{subsubsect:mershearwaves}
The evolution of the internal rotation profile changes drastically when IGW are taken into account in conjunction with meridional circulation and shear turbulence, as can be seen in Fig.~\ref{fig:omegaHzevol1p0} (top plots).
As already discussed in \S~\ref{sec:IGWexcitation}, the emitted wave spectrum strongly evolves with the stellar structure along the PMS. 
IGW are first emitted by the receding convective envelope, and much later by the convective core when it appears during the final approach towards the zams. 
In the case of the 1~M$_{\odot}$ model, IGW emitted by the convective core play actually no role since their luminosity is extremely low (see Fig.~\ref{fig:excitrenv} and discussion in \S~\ref{sec:IGWexcitation}).
Therefore the following discussion refers only to those emitted by the envelope.

\begin{figure*}
\includegraphics[angle=0,width=8.5cm]{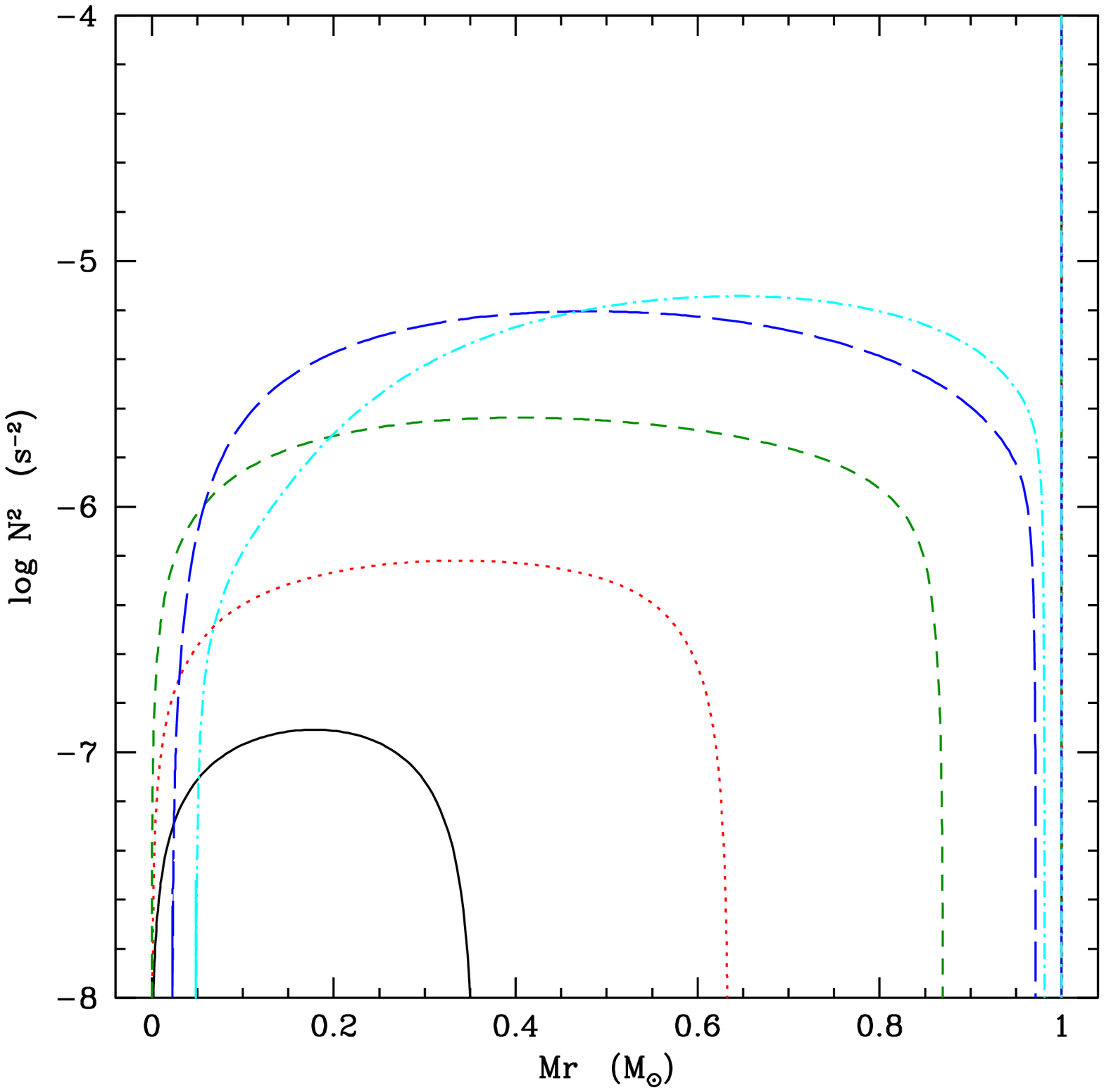} 
\includegraphics[angle=0,width=8.5cm]{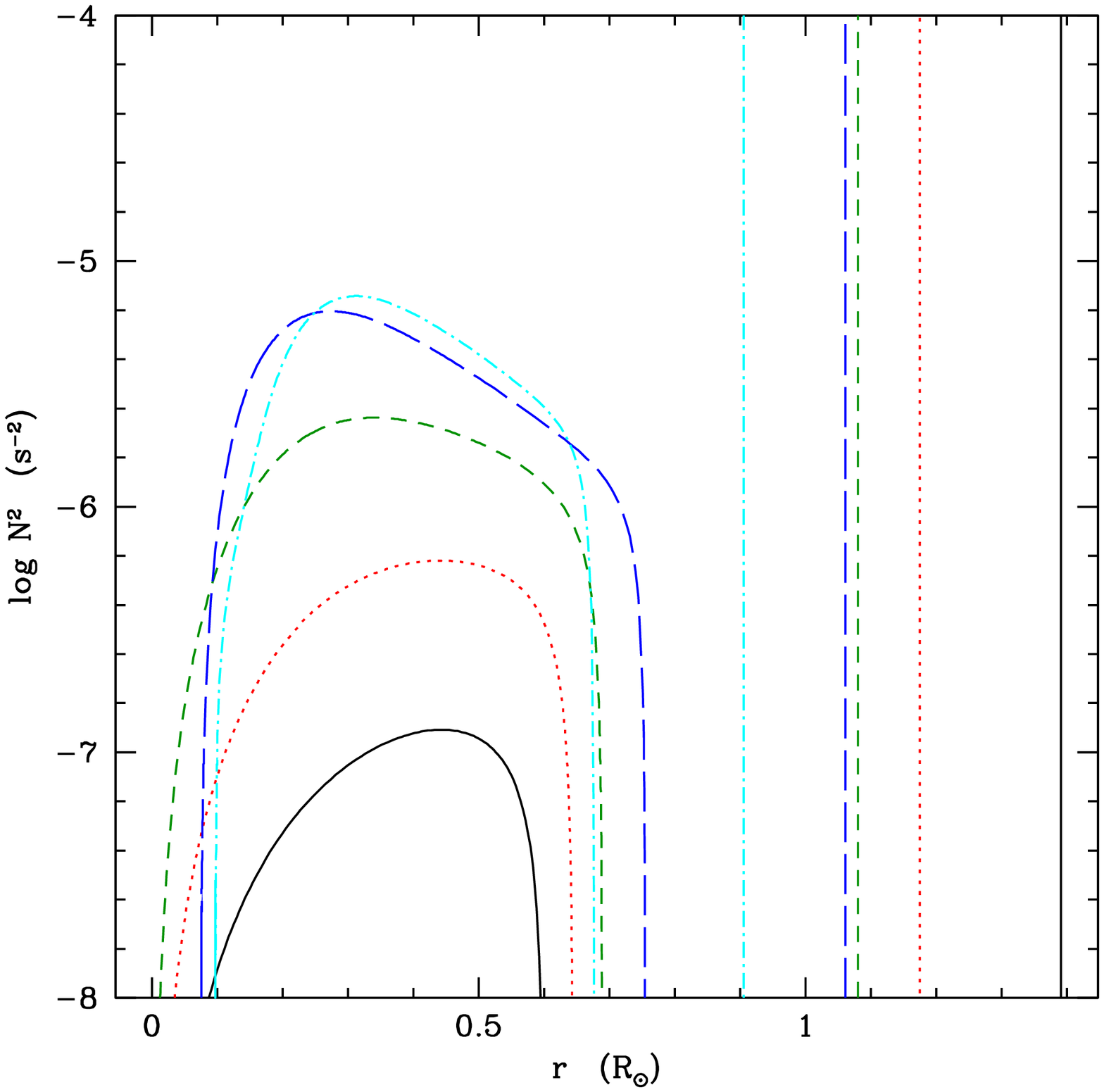}
\caption{Evolution of the Brunt-V\"ais\"al\"a frequency in the 1~M$_{\odot}$ PMS star as a function of relative mass fraction and radius (left and right respectively). The colours correspond to the same ages as in Fig.~\ref{fig:omegaHzevol1p0}. On the right plot the vertical lines indicate the total stellar radius at the corresponding ages}
\label{fig:BVf}
\end{figure*}

In order to understand wave-induced transport, we must also focus on the important quantities for wave damping in the radiative layers, 
namely the Brunt-V\"ais\"al\"a frequency $N^2$ and the thermal diffusivity $K_T$: For a given differential rotation within the radiative layers, low-frequency (i.e., with $\omega < $ 3.5~$\mu$Hz) and/or large degree waves that dominate the angular momentum transport are damped very efficiently close to the convective edges when $N^2_T$ is too small or when $K_T$ is too large (see Eq.~\ref{optdepth}).
Fig.\ref{fig:BVf} and \ref{fig:chemicaldiffusioncoefficients} show the radial profiles of these two quantities in the radiative layers of the 1~M$_{\odot}$ model at various ages (see also Fig.~\ref{fig:surfdetails} and \ref{fig:coredetails} that show the variations along the evolution track of the value of $K_T$ just below the convective envelope and above the convective core).
At all ages $N^2$ drops near the stellar center and the 
convective edges; in addition its value at a given depth increases with time along the PMS 
as a result of the stellar contraction that leads to an increase of gravity and a decrease of the pressure scale height as the star evolves.
On the other hand the value of $K_T$ just below the convective envelope also increases as the star contracts and move towards higher effective temperature; this implies stronger damping of all the waves (independently of their properties) closer to the convective envelope; note that $K_T$ at a given depth within the star increases only slightly during the evolution.  Besides, the build up of differential rotation with time within the star induces a change in the local Doppler shift frequency, which allows a different damping for waves with different frequencies and $m$ through the term
$\sigma^{-4}  \left({N^2 - \sigma^2}\right)^{-0.5}$. 

Let us see what these general considerations imply for the 1.0~M$_{\odot}$ model. 
We start with initial solid body rotation and then follow the transport of angular momentum when the radiative layers appear.
At that moment differential rotation has not yet developed, and the local frequency $\sigma$ of individual waves in the very thin radiative zone is similar to their emission frequency $\omega$ at the base of the convective envelope. However slight differential rotation soon builds up as a result of stellar contraction along the Hayashi track, which induces a Doppler shift between the emission and local IGW spectra. 
As a consequence, low-frequency low-degree waves, which undergo the largest differential damping between retrogade and prograde components, soon penetrate all the way to the central regions where they deposit their negative momentum and very efficiently spin down the core whose amount of angular momentum is minute (see Fig.~\ref{fig:omegaHzevol1p0}).
This explains the strong positive gradient in the profile of $\Omega$ below $\sim 0.2$ R$_{\odot}$, while the negative gradient of $\Omega$ in the external layers results from ongoing stellar contraction. 
As a consequence a peak builds up in the internal rotation profile with a core spinning at lower rate than the stellar surface all along the PMS. 

We show in Fig.\ref{flux} the total flux of angular momentum carried by the waves as a function of depth within the 1~M$_{\odot}$ model, and compare it to the contribution of meridional circulation and shear turbulence at different evolution stages. 
We note first that the transport of angular momentum is generally dominated by the waves in the radiative layers where they can propagate, except in the early times when 
meridional circulation dominates in the most external regions below the convective envelope (upper panel at 14~Myr). 
Since downward propagating waves are totally damped as soon as the condition $ \Omega(r) = \omega /m +\Omega _{\rm cz}$
is fulfilled near the corotation radius, the total flux F$_{IGW}$ drops and remains negligible below the  $\Omega$ peak. This can be clearly seen in the middle and lower panels in Fig.\ref{flux} at 35 and 55~Myrs; at that time meridional circulation dominates in the regions below $\sim$ 0.15 and 0.2~M$_{\odot}$ respectively, while IGW are dominant in the outer regions. 
Note that the total flux of angular momentum is dominated by IGW when they are accounted for and is larger by two orders of magnitude compared to the case without IGW.
Overall, IGW do shape the circulation patterns, leading to the appearance of several loops in the whole radiative region as can be seen in Figs.~\ref{flux} and \ref{fig:U2}. 

Let us add a final remark. 
As explained in \S~\ref{subsubsection:generaltransport}, we have increased by a factor 2 the IGW luminosity in order to account for the results by \citet{LecoanetQuataert13} who predict the IGW flux due to turbulent convection to be a few to five times larger than in previous estimates by e.g. \citet{GoldreichKumar90} and \citet{GoldreichMurray1994}. 
In order to test the impact of this assumption, we have computed two additional models for the 1~M$_{\odot}$ rotating star with multiplying factors of 1 and 5. 
We find that this has no impact on the conclusions, as can be seen in Fig.~\ref{fig:allzams} where we plot the corresponding rotation profiles at the arrival on the ZAMS. 

\subsection{Impact of the stellar mass}
\label{subsection:variousmasses}

\begin{figure}
  \includegraphics[width=0.5\textwidth]{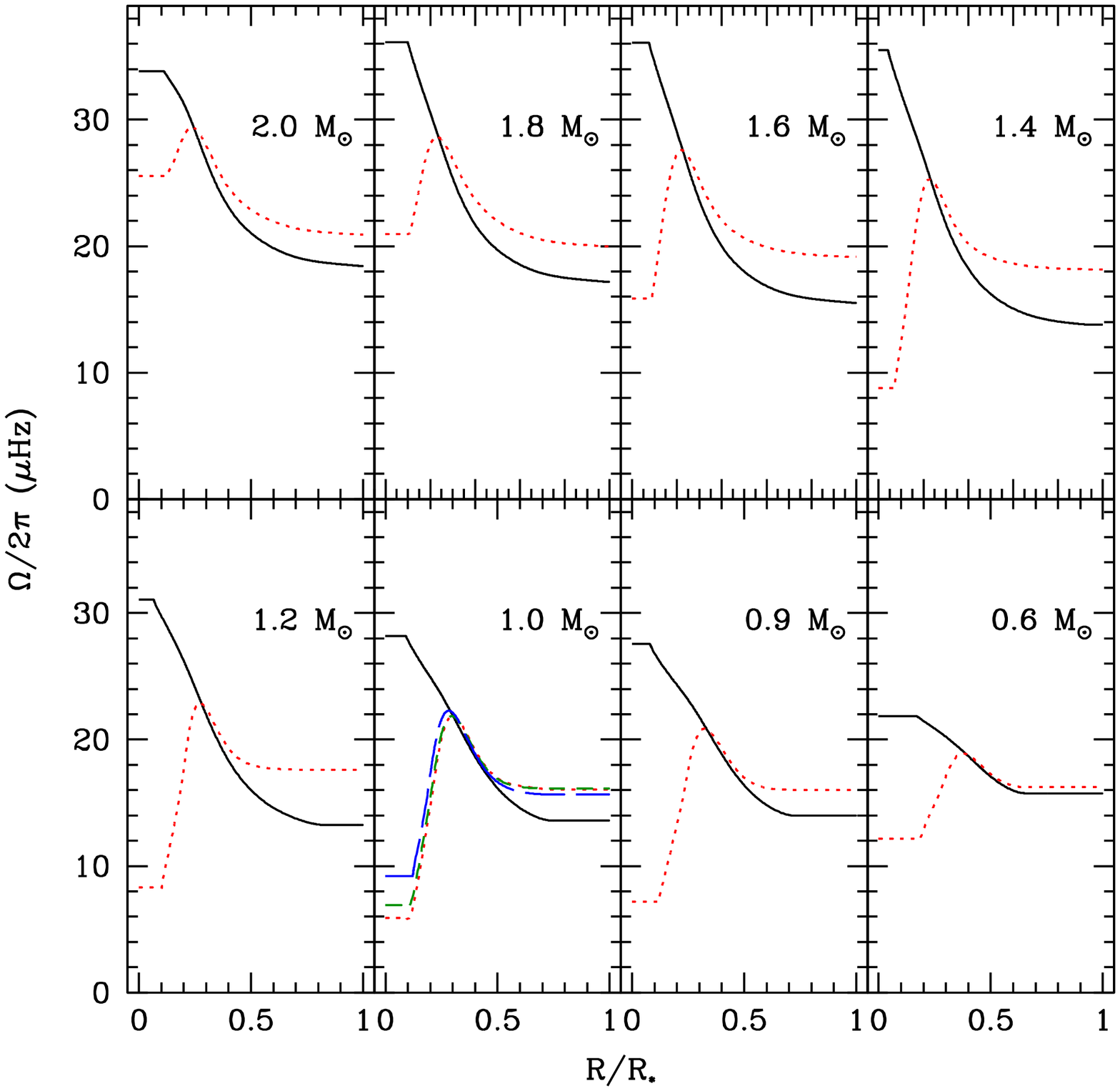}
  \caption{Rotation profile on the ZAMS 
for all the stellar masses between 0.6 and 2.0~\Ms{} in the cases without and with IGW (full black and red dotted lines respectively). 
In the 1~M$_{\odot}$ panel, the blue long-dashed and the green dashed lines correspond to computations made with multiplication factors for IGW luminosity 	
of one and five respectively, all the other models with IGW being computed with a multiplication factor of two (see Eq.~\ref{ev_omega})
}
  \label{fig:allzams}
\end{figure}

For all the stars within the considered mass range, strong differential rotation with a fast rotating core is obtained under the combined action of stellar contraction and meridional circulation 
when IGW are not accounted for. 
Besides, in all cases IGW do break-up the stellar core, which results in a peak in $\Omega$ at r$\sim$0.25-0.3R$_*$ as in the 1~M$_{\odot}$ case. 
This can be seen in Fig.~\ref{fig:allzams} where we show the rotation profiles at the arrival on the ZAMS for all our models (black and red lines correspond respectively to the models computed without or with IGW).

\begin{figure}
  \includegraphics[width=0.5\textwidth]{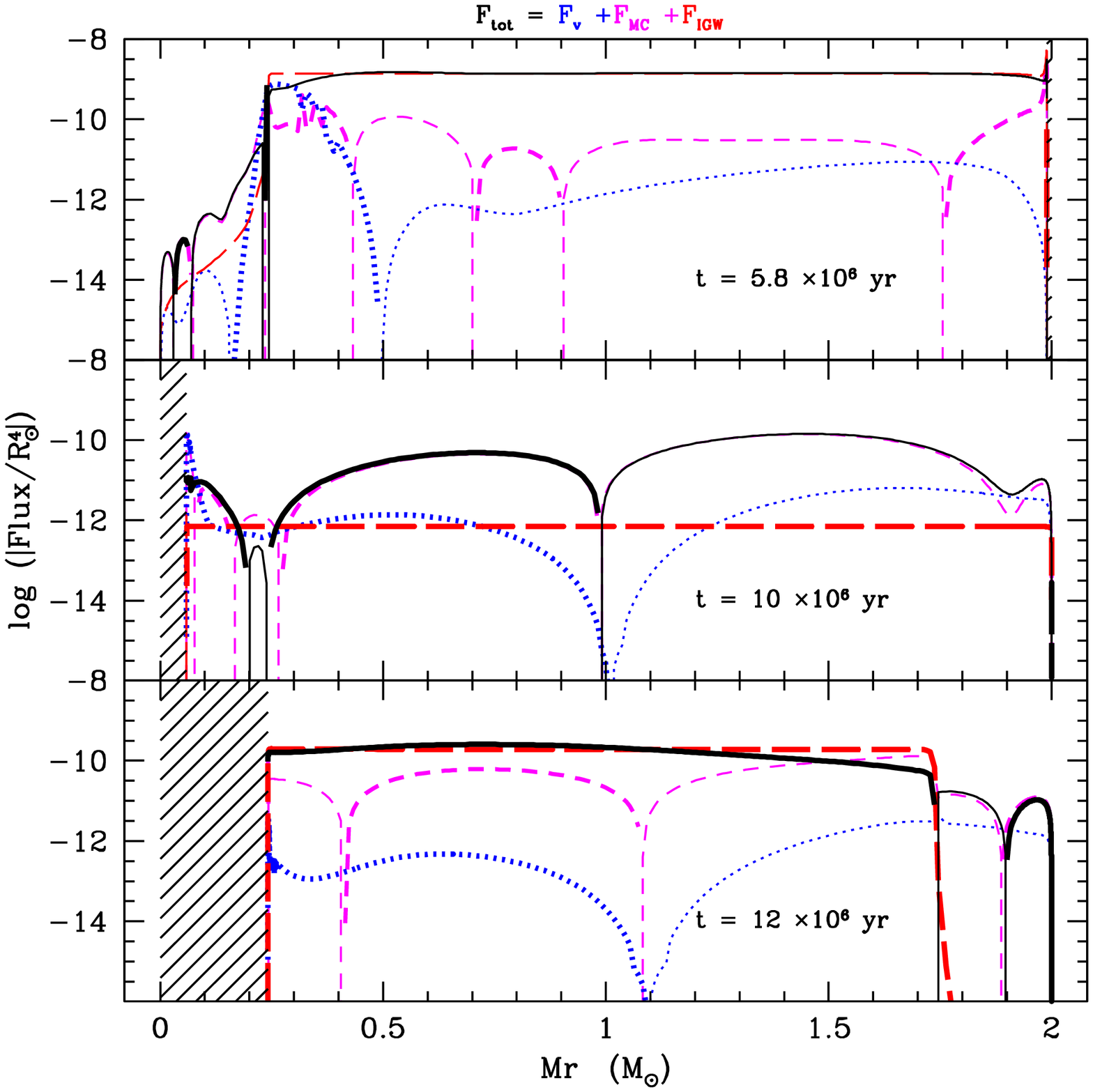}
    \caption{Same as Fig.~\ref{flux} for the 2~M$_{\odot}$ model computed with IGW}
  \label{fig:2mfluxam}
\end{figure}

\begin{figure*}
  \includegraphics[width=0.45\textwidth]{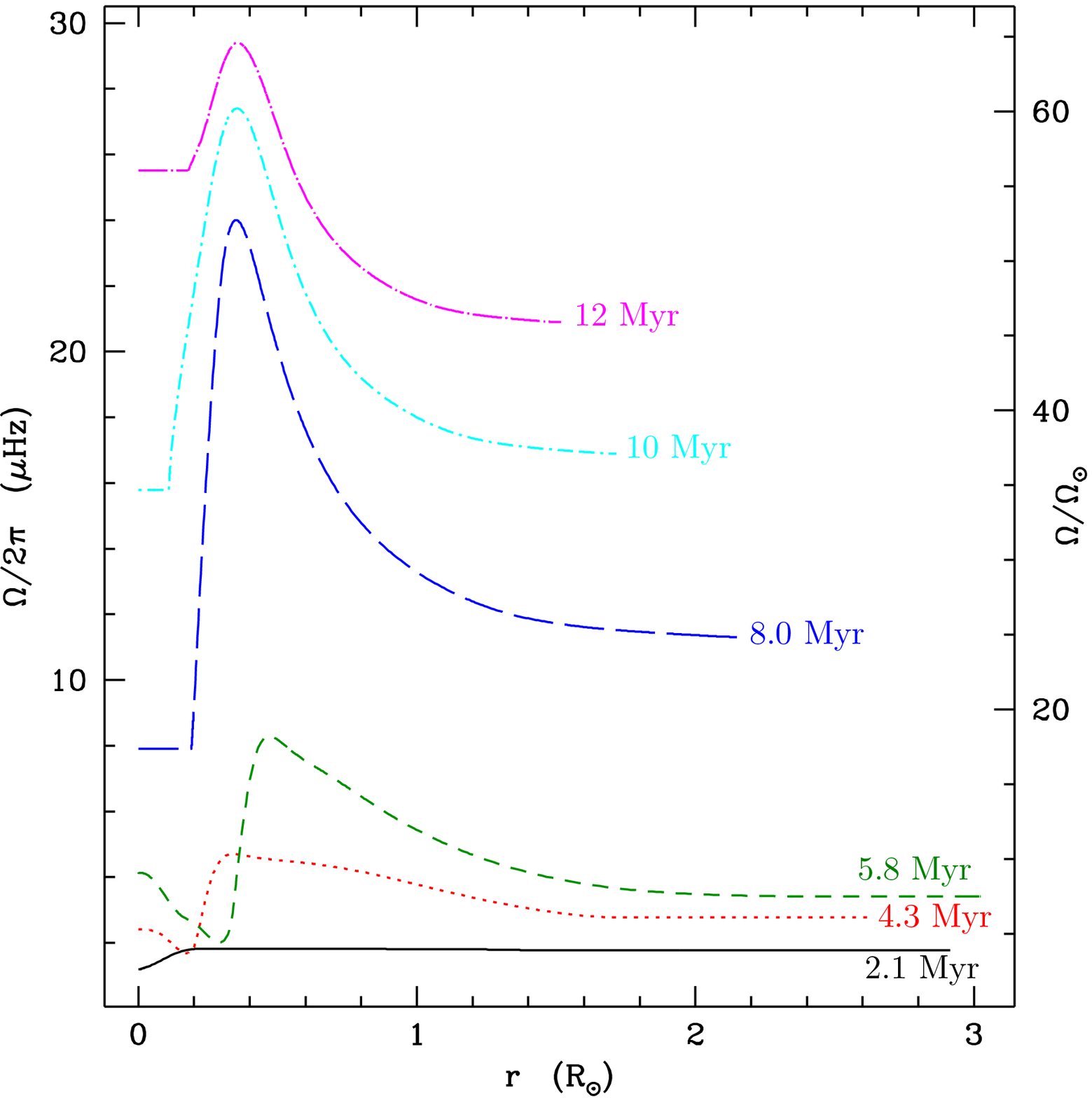}
  \includegraphics[width=0.45\textwidth]{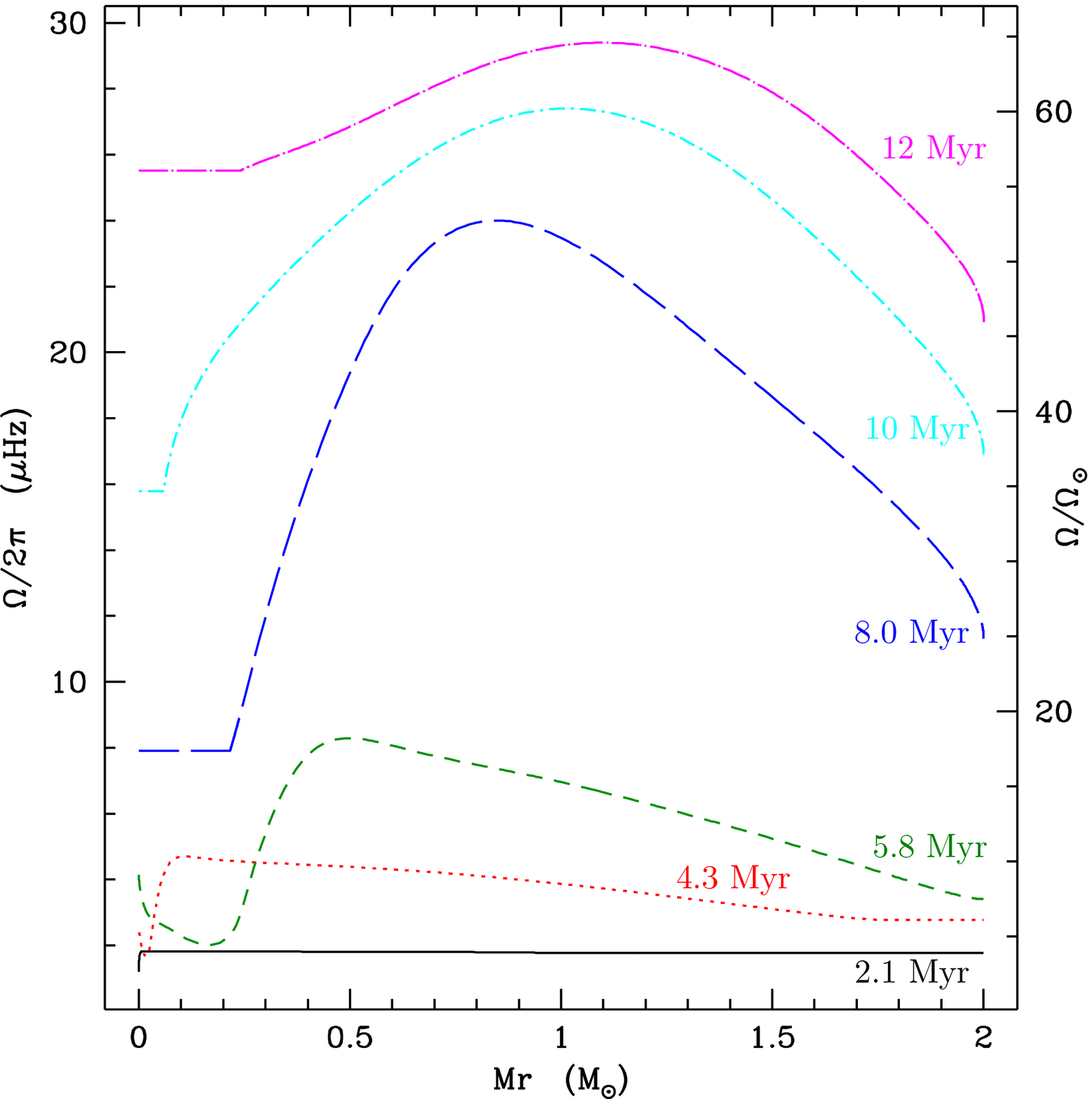}
  \caption{Same as Fig.\ref{fig:omegaHzevol1p0} for the 2~M$_{\odot}$ model computed with IGW
}
  \label{fig:2msunomega}
\end{figure*}

Let us note however that the impact of IGW is slightly different in stars more massive than $\sim 1.6$~M$_{\odot}$. 
This is illustrated for the 2~M$_{\odot}$ star in Fig.~\ref{fig:2mfluxam} where we decompose the total flux of angular momentum within the model according to the various transport processes at three different ages, 
and in Fig.~\ref{fig:2msunomega} where we follow the corresponding evolution of the radial profile of $\Omega$.
For this more massive star, IGW emitted by the convective envelope dominate during the first part of the PMS and manage to slow down the most central regions as in the 1~M$_{\odot}$ case (top panel, Fig.~\ref{fig:2mfluxam}).
However those waves fade away when the convective envelope becomes too thin and are supplanted by those emitted by the convective core at the approach of the ZAMS (see Fig.~\ref{fig:igwexcit}). 
During that transition period (middle panel in Fig.~\ref{fig:2mfluxam}), meridional circulation dominates the transport of angular momentum although shear turbulence also contributes more efficiently near the most central regions (between 0.05 and 0.1~\Ms{}) and in the most external layers; as a result, the core slightly accelerates and eventually manages to rotate faster than the outer radiative layers, although not fast enough for the peak to be erased. 
Once the convective core has sufficiently developed (lower panel, Fig.~\ref{fig:2mfluxam}), the IGW emitted in the central regions will start conveying angular momentum very efficiently towards the core; 
at that time meridional circulation remains however the dominant process in the most external radiative layers. 

\section{Global stellar properties, surface rotation and lithium abundance}
\label{section:globalproperties}

\begin{figure*}
  \includegraphics[width=0.5\textwidth]{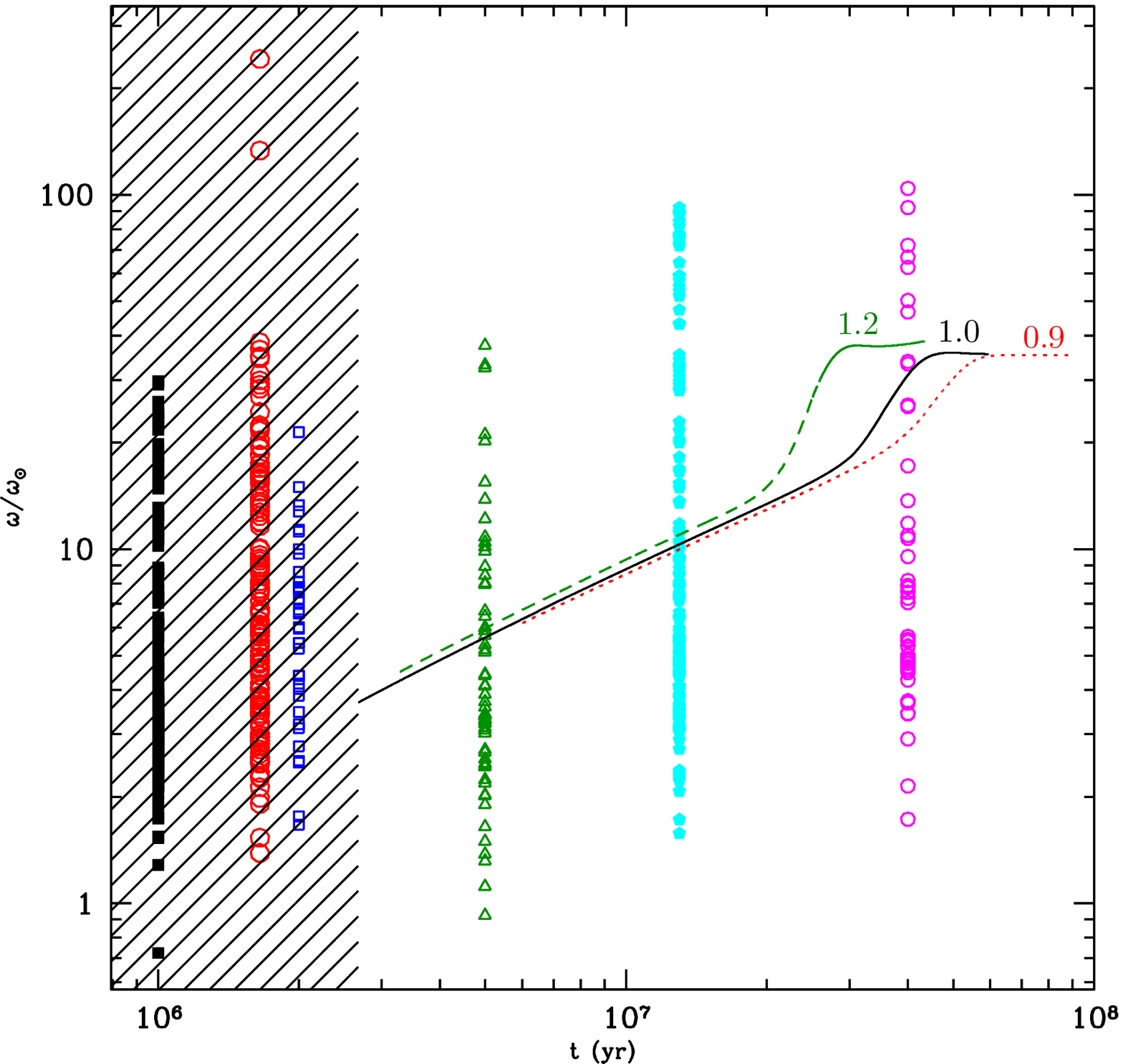}
  \includegraphics[width=0.5\textwidth]{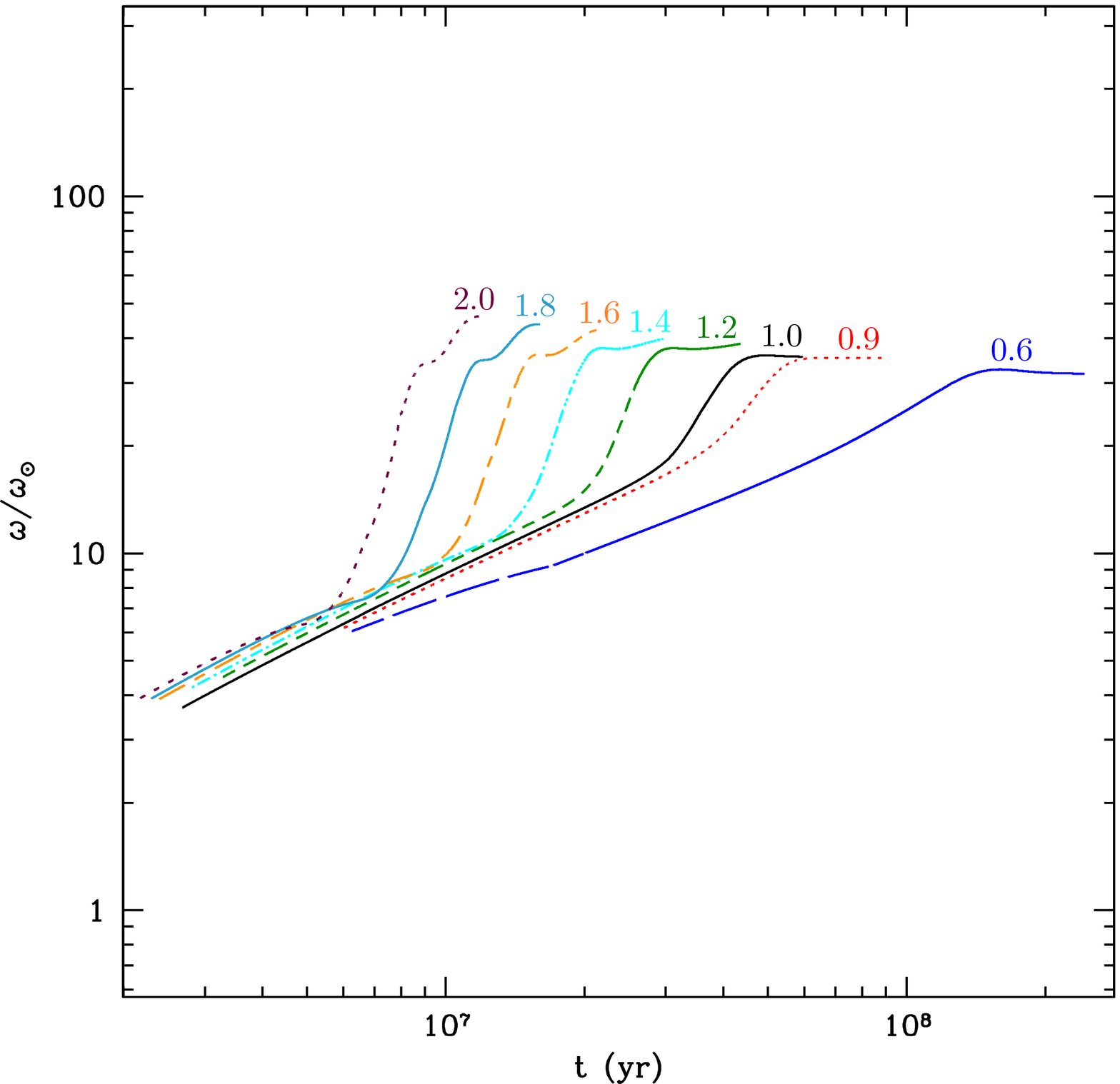}
  \caption{Evolution of the surface rotation rate as a function of time for the models computed with IGW. Stellar masses are indicated on the tracks. 
  In the left panel, the theoretical predictions for the 0.9, 1, and 1.2~M$_{\odot}$ models are compared with 
  the observed rotational distribution for stars with estimated masses between $\sim$ 0.9 and 1.1~M$_{\odot}$ in young open clusters from \citet{GalletBouvier13}
}
  \label{fig:surfacerotation}
\end{figure*}

\begin{figure*}[t]
\includegraphics[angle=0,width=8.5cm]{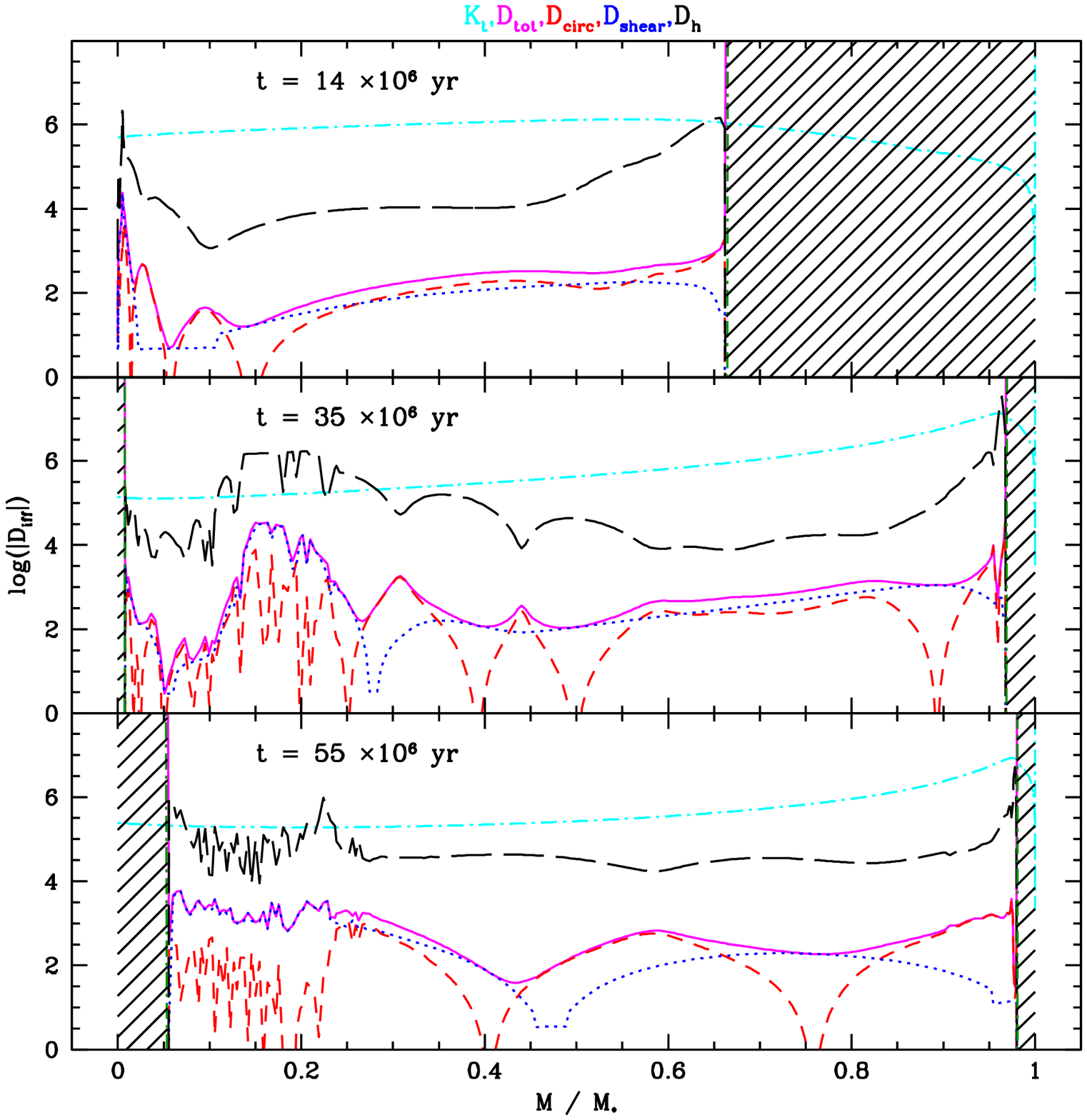}
\includegraphics[angle=0,width=8.5cm]{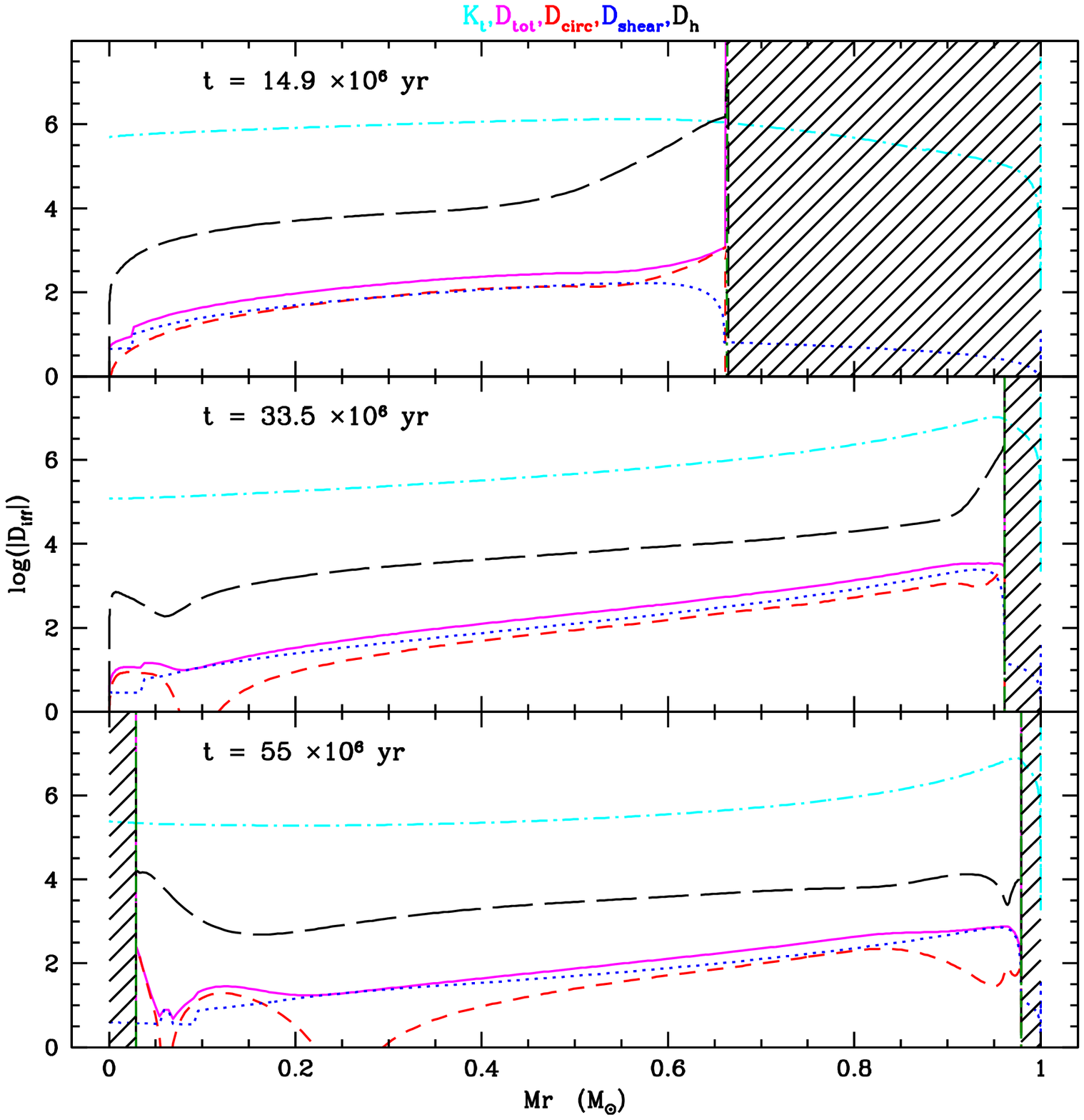}
\caption{Diffusion coefficients associated to meridional circulation (red), horizontal and vertical turbulence (black and blue respectively). The total diffusion coefficient for the chemicals (magenta) and thermal diffusivity (cyan) are also shown.  
The figures correspond to the 1~M$_{\odot}$ models with and without IGW (left and right respectively) at different evolution ages along the PMS.  Hatched areas indicate the convective regions}
\label{fig:chemicaldiffusioncoefficients}
\end{figure*}

We summarize in Table~\ref{table:properties} the main properties of our models computed under various assumptions. 
We also include the predictions for two additional models of 1~M$_{\odot}$ that account for the hydrostatic effects of rotation (i.e., the effects of centrifugal acceleration on effective gravity) and show in 
Fig.~15 all the corresponding evolution tracks for this star. 
We see that the rotating tracks without hydrostatic effects are hardly modified compared to the standard case, the main shift to slightly lower effective temperature and luminosity (that implies slightly longer PMS lifetime) being due to the effects of the centrifugal force and not to rotation-induced mixing. This is in agreement with the predictions by \citet{Eggenbergeretal12} (see also \citealt{Pinsonneaultetal89,MartinClaret96,Mendesetal99}).  
However the hydrostatic effects are modest and our general conclusions on the evolution of the internal rotation profile and on the impact of IGW are not affected by this simplification.
We can also note in Table~\ref{table:properties} that the models computed with IGW have longer PMS lifetimes than the others. 
This simply results from the higher total diffusion coefficient for chemicals in the deep radiative layers close to the convective core when central H-burning sets in close to the ZAMS (see Fig.~\ref{fig:chemicaldiffusioncoefficients}).
 
As shown in Fig.~\ref{fig:surfacerotation}, the evolution of surface rotation for the models with IGW accounts well for the mean rotation rates collected by \citet{GalletBouvier13} 
for PMS stars in young open clusters in the considered mass range.
The rotation velocity at the arrival on the ZAMS is slightly higher (by a few $\%$; see Table~\ref{table:properties}) in this case than in rotating models without IGW, due to the different efficiency of the redistribution of angular momentum by the various transport mechanisms within the star as discussed previously. Again, the hydrostatic effects are negligible.

\begin{figure}[t]
\includegraphics[angle=0,width=8.5cm]{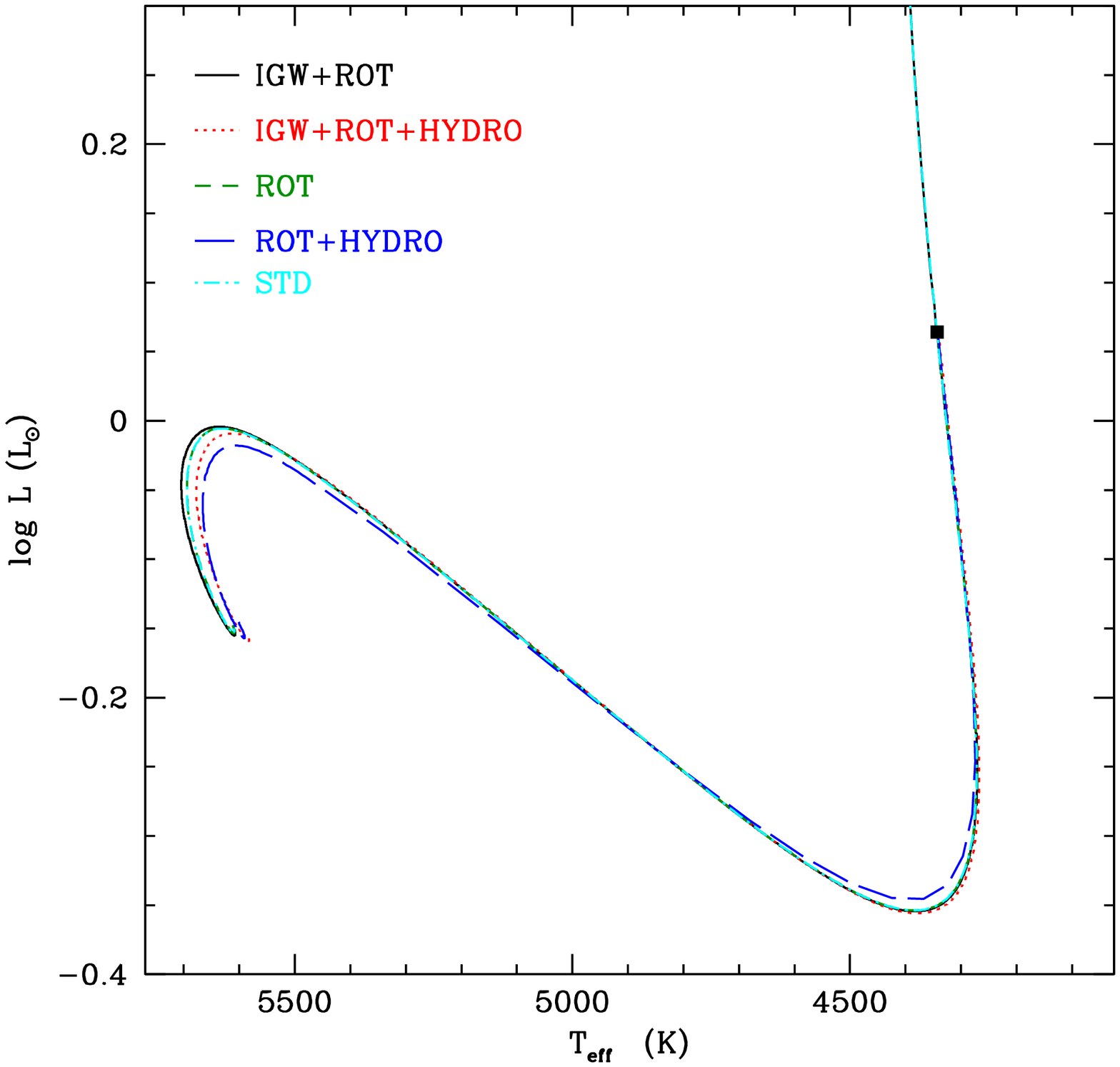}
\caption{Impact of rotation, IGW, and of the hydrostatic effects on the evolution track of  the 1~M$_{\odot}$ star. The square indicates the point where the radiative zone appears and internal transport of angular momentum starts}
\label{fig:hrd1Msun_all}
\end{figure}

The surface lithium abundance at the ZAMS is not significantly different in the rotating models without and with IGW, as can be seen from Table~\ref{table:properties}. 
Indeed this quantity mostly depends, on one hand, on the temperature at the base of the convective envelope, which is unaffected since the evolution tracks almost superpose, and on the other hand, on the diffusion coefficient $D_{eff}$ (Eq.~\ref{eq:Deff}) in the external radiative layers shown in Fig.~\ref{fig:chemicaldiffusioncoefficients}. 
Since the gradient of $\Omega$ in the outer part of the star is dominated by stellar contraction and is very similar in the cases with and without IGW (see Figs.~\ref{fig:omegaHzevol1p0} and \ref{fig:2msunomega}), the resulting Li abundance at the ZAMS is unaffected.
The rotating models including the hydrostatic effects have slightly higher lithium abundance on the ZAMS, in agreement with the behavior found by \citet{Eggenbergeretal12}.
In a future work we will revisit PMS Li depletion taking the influence of the disk lifetime, of the initial rotation velocity, and of magnetic braking into account . 

\section{Conclusions}
\label{sec:conclusion}

In this paper we have analyzed the transport of angular momentum during the PMS for solar-metallicity, low-mass stars (with masses between 0.6 and 2.0~M$_{\odot}$) through
the combined action of structural changes, meridional circulation, shear turbulence, and internal gravity waves generated by Reynold-stress and buoyancy in the stellar convective envelope and core (when present). 

For all the stellar masses considered, IGW are efficiently generated by the convective envelope with a momentum luminosity that peaks around $T_{eff} \sim$6200~K, as in the case of main sequence stars. 
These waves soon become an efficient agent for angular momentum redistribution because they spin down the stellar core early on the PMS, while structural changes lead to a negative differential rotation in the outer stellar layers as the star contracts.
On the other hand, IGW generated by the convective core close to the arrival on the ZAMS carry much less energy, except in the case of stars more massive than $\sim$ 1.6~M$_{\odot}$.
Over the whole considered mass range, IGW were found to significantly modify the internal rotation profile of PMS stars and lead to slightly higher surface rotation velocity compared to the case where only meridional circulation and shear turbulence are accounted for. 

The exploratory results presented in this paper show the ability of IGW to efficiently extract angular momentum in the early phases of stellar evolution, as anticipated by \citet{TalonCharbonnel2008} and as shown by \citet{CharbonnelTalon2005Science} and \citet{TalonCharbonnel2005} for solar-type main sequence stars.
We now plan to investigate the influence of the disk lifetime, of the initial rotation velocity, and of magnetic braking during the PMS over a broader mass domain in order to compare model predictions with large data sets that are currently being collected to trace the rotational properties of young stars.  
 
\begin{acknowledgements}
We thank F.Gallet and J.Bouvier for kindly providing data before publication and for fruitful discussions, as well as P.Eggenberger for detailed model comparisons. We thank the referee J.P.Zahn for suggestions that helped improve the manuscript.
We acknowledge financial support from the Swiss National Science Foundation (FNS), from the french Programme National de Physique Stellaire (PNPS) of CNRS/INSU, and from the Agence Nationale de la Recherche (ANR) for the project TOUPIES (Towards Understanding the sPIn Evolution of Stars).
\end{acknowledgements}

\bibliographystyle{aa}
\bibliography{BibADS}

\end{document}